\documentclass[conference]{IEEEtran}
\usepackage[LGR,T1]{fontenc}
\usepackage{lmodern}

\usepackage[latin9]{inputenc}
\usepackage{array}
\usepackage{prettyref}
\usepackage{float}
\usepackage{textcomp}
\usepackage{pdfpages}
\usepackage{multirow}
\usepackage{amstext}
\usepackage{graphicx}

\usepackage{listings}
\usepackage{color}

\definecolor{dkgreen}{rgb}{0,0.6,0}
\definecolor{gray}{rgb}{0.5,0.5,0.5}
\definecolor{mauve}{rgb}{0.58,0,0.82}

\usepackage{babel}
\usepackage{fancyhdr}
\usepackage{url}

\usepackage[pdfpagelabels]{hyperref}

\usepackage{tabulary}

\DeclareRobustCommand{\greektext}{%
  \fontencoding{LGR}\selectfont\def\encodingdefault{LGR}}
\DeclareRobustCommand{\textgreek}[1]{\leavevmode{\greektext #1}}
\ProvideTextCommand{\~}{LGR}[1]{\char126#1}

\begin{document}

\hypersetup{pageanchor=false}

\title{A Survey on Property-Preserving Database Encryption Techniques in the Cloud}
%\author{Johannes Koppenwallner \\
%	University of Vienna \\
%    Email: a9405411@unet.univie.ac.at
%	\and
%	Erich Schikuta \\
%	University of Vienna \\
%    Email: erich.schikuta@univie.ac.at\\
%	}

\author{\IEEEauthorblockN{Johannes Koppenwallner}
\IEEEauthorblockA{\textit{Faculty of Computer Science} \\
\textit{University of Vienna}\\
Vienna, Austria}
\and
\IEEEauthorblockN{Erich Schikuta}
\IEEEauthorblockA{\textit{Faculty of Computer Science} \\
\textit{University of Vienna}\\
Vienna, Austria \\
erich.schikuta@univie.ac.at}
}

\maketitle

\begin{abstract}
Outsourcing a relational database to the cloud offers several benefits,
including scalability, availability, and cost-effectiveness. However,
there are concerns about the security and confidentiality of the outsourced
data. A general approach here would be to encrypt the data with a
standardized encryption algorithm and then store the data only encrypted
in the cloud. The problem with this approach, however, is that with
encryption, important properties of the data such as sorting, format
or comparability, which are essential for the functioning of database
queries, are lost. One solution to this problem is the use of encryption
algorithms, which also preserve these properties in the encrypted
data, thus enabling queries to encrypted data. These algorithms range
from simple algorithms like Caesar encryption to secure algorithms
like mOPE. The report at hand presents a survey on common encryption techniques
used for storing data in relation Cloud database services. It presents the applied methods
and identifies their characteristics.
\end{abstract}

\begin{IEEEkeywords}
Database systems, Property Preserving encryption, Cloud computing
\end{IEEEkeywords}

%\pagebreak{}

%\tableofcontents{}\pagebreak{}

\pagenumbering{arabic}\pagestyle{fancy}

\section{Introduction}

\subsection{Motivation}

Operating and maintaining a database is a laborious task. At first
hardware for the server is needed. This hardware is often over powered
for the expected, but under powered for the peak load. Then this hardware
needs space, has to be properly maintained and secured against failure.
The operating system has to be administered and updated regularly
and the database management system also has to be installed and maintained.
If the use of a commercial database is needed, complying to the licenses
is a challenge for itself. In short, operating a database does not
only require a significant investment, but also permanent effort (staff
and means) for its maintenance. If requirements for the database change
at some later point, for example the expected usage is much higher
than predicted, adjusting the solution can be expensive. Here is where
the cloud comes into play. In the database as a service model, hardware
and software are completely maintained by the cloud vendor. No upfront
investment is needed, and if requirements change the service model
can also be easily adapted. But there is a significant drawback for
outsourcing a database into the cloud: The control over the data is
lost. Before putting the data into the cloud, only members (for example
administrators) of the organization controlling the data were involved,
now as second party, the cloud vendor is involved, too. This is the
point, where many organizations give up their plan of moving their
data into the cloud. It is simply not worth the risk and the additional
organizational effort. An apparent easy solution to this problem would
be to encrypt all data in the cloud with a standard cipher. Doing
this, the cloud vendor would never have access to the plain text,
so there should be no more issues regarding privacy and confidentiality.
The problem with this solution is, that it does not work as intended.
By using standard ciphers for encryption, the relational data model
is not usable anymore. The reason for this is that encrypted values
are no more compatible with the defined column data types, and queries
do not retrieve the correct results anymore, because important properties
of the plain text, like identity or order are lost. So we have two
solutions with different drawbacks: Deploying the data unencrypted
is insecure and usable, while encrypting the data makes it secure,
but unusable. This report tries to elaborate and show alternative
solutions, which maintain most of the security, while still being
usable.

%Our research in this area is based on our experience with large scale data stores~\cite{schikuta1998vipios} and complex applications in Grids and Clouds~\cite{mach2012generic,schikuta2004n2grid,cs745}, and strongly motivated by our focus on Web-based workflow optimizations~\cite{schikuta2008grid,kofler2009parallel} and their respective management~\cite{stuermer2009building}.

\subsection{Overview}

For storing and retrieving structured data, the relational data model
is still the dominant model. More and more data is collected and stored
in databases and they are a critical part in nearly every IT environment.
Traditionally these databases are run in house and managed by members
of the same organization using it. With the rise of cloud computing
this is changing. Databases are outsourced into the cloud and run
and managed by the cloud service provider. As mentioned, this leads
to serious privacy and security concerns, because not only the members
of the organization itself have access to the data, but additionally
the administrators of the cloud provider have access. Another serious
concern is that a database, which was formally only accessible in
an internal network, is now accessible over the internet. A solution
to this problem is to encrypt the data with proven secure ciphers
before putting it into the cloud. This approach does not work with
structured data, because important properties of the data are lost
during encryption. The result is that the relational model does not
work for the encrypted data anymore. The format of the data has changed
and queries do not work the way they used to on the plaintext data.
The data model and any application depending on this schema have to
be changed. Even then, the result comes with a serious performance
penalty, which makes this approach often impractical. To avoid this,
other solutions are required. A lot would be gained, if ciphers can
encrypt the values while still keeping format, order or other query
relevant properties. In the optimal case, the data model can be left
unchanged, while still providing data confidentially by encryption.
Of course, any application depending on such an unchanged data model
can be left unchanged too, if encryption and decryption is done transparently.
To achieve the objective of a fully usable encrypted relational database
in the cloud, multiple problems have to be solved.

\paragraph*{Cryptography}

A short overview of cryptography is given. This includes history,
taxonomy and the description of some of the most significant ciphers.
Ancient ciphers like Caesar's and standard ciphers like DES and AES
are shown. A small example of the classical Caesar's cipher is presented.
The next chapter describes some attack scenarios and the use of encryption
to mitigate these threats in the context of a database. It shows the
use case for data at rest and data in transit. For data at rest it
shows the different levels (storage, database, application) at which
encryption can be performed. The advantages and disadvantages of the
place of encryption are discussed here too. Then concrete solutions
and applications of encryption on the different levels are presented.
Database specific issues of ciphers are shown, and state of the art
encryption techniques, such as homomorphic and order preserving encryption
are described.

\paragraph*{Relational Model Requirements}

The relational model has some implicit requirements which have to
be satisfied to be usable. As plain text always satisfies these requirements,
ciphertext does often not (at least not out of the box). As these
requirements are different for each SQL construct, the specific requirements
for data definition and queries are given.

\paragraph*{Ciphers with Properties}

Although standard ciphers often do not satisfy the properties required
by the relational model, ciphers exist which satisfy some or multiple
of these requirements. An overview of the different properties like
format preserving, order preserving, functional and homomorphic encryption
is given. State of the art ciphers are shown and described in detail.
For queries, order preserving is a often needed property. Different
order preserving encryption schemes are shown in detail. Security
definitions are given, and theses ciphers are compared regarding their
properties, security and implementation.

\paragraph*{Commercial Solutions}

An overview of existing solutions is given. These solutions includes
research projects like CryptDB as well as commercial products and
solutions. These cloud based solutions are presented and compared
with each other regarding their features and security.

\paragraph*{Requirements for Cloud Computing}

Moving data to the cloud requires additional concerns especially,
but not only regarding security. After describing the database as
a service scenario based on the previous chapter, additional requirements
on security and privacy in the cloud are given. Contradictions between
security requirements and other cloud-specific requirements, like
scalability or elasticity, are shown. Deployment scenarios are described,
and a short overview of available database as a service solutions
is given. The chapter is concluded with the description of Relational
Cloud, a project aiming to enhance the existing DBaaS model with security
and privacy in focus.

\subsection{Scope and Limitations}

This report focuses on relational databases running in the the cloud.
For evaluation the database as a service (DBaaS) model is used from
some cloud vendors. The encryption of unstructured data or data in
non-relational databases like NoSQL databases is not examined. Of
course, some of the presented ciphers may work here as well.
This report is focused on the encryption of relational databases in
the cloud, not on security in general. Other important security topics
like key management, authorization, authentication are only scratched
on the surface or not discussed at all.

%\newpage{}

\section{Cloud}

Originating from data centers and grid computing, cloud computing
started gaining momentum around 2006. Organizations and enterprises
began to outsource part of their internal IT into the cloud, while
a few other companies provided their internal services to external
customers and became cloud service provider. One of the first of theses
companies was Amazon starting their cloud offer ``Amazon Web Service''
(AWS) in 2006. Google and Microsoft followed later with their offers
``Google Cloud Platform'' and ``Windows Azure'' (later renamed
to Microsoft Azure). Another kind of company here to mention is Salesforce.
This company provided its software as a service over the internet
from the start, thus becoming one of the first software as a service
provider. As the definition of cloud computing is is still evolving
and changing~\cite{Mell:2011:SND:2206223}, for this work the definition
of cloud computing from the National Institute of Standards (NIST)
is used:
\begin{quote}
``Cloud computing is a model for enabling ubiquitous, convenient,
on-demand network access to a shared pool of configurable computing
resources (e.g., networks, servers, storage, applications, and services)
that can be rapidly provisioned and released with minimal management
effort or service provider interaction''~\cite[pp. 2]{Mell:2011:SND:2206223}.
\end{quote}
Focusing more on the business perspective, but otherwise similar is
another definition of cloud computing: ``Cloud computing is an IT
deployment model, based on virtualization, where resources, in terms
of infrastructure, applications and data are deployed via the internet
as a distributed service by one or several service providers. These
services are scalable on demand and can be priced on a pay-per-use
basis''~\cite[pp 4]{LeimeisterEtAl:2010}. According to~\cite{LeimeisterEtAl:2010},
there is a strong trend from products to services. This is not only
true for hardware, but more and more for software, too. In cloud computing,
hardware and software as a service are tightly integrated. Multiple
services can be composed to support complex business processes. The
NIST~\cite{Mell:2011:SND:2206223} defined the essential characteristics
of cloud computing as:
\begin{itemize}
\item On-demand self-service.
\item Broad (ubiquitous) network access.
\item Resource pooling.
\item Rapid elasticity.
\item Measured service.
\end{itemize}
On-demand self-service means that a customer can add or remove computing
capabilities without any human interaction as wanted. All capabilities
and services are provided and accessed over a network. The provider's
computing resources are pooled to serve multiple clients. Another
characteristic is that resources can be elastically provisioned and
released. The last characteristic is that resource usage is monitored,
controlled and reported. All these characteristics are fulfilled by
the services from the established cloud vendors. As an example Amazon
Web Services (AWS) allows customers to manage their cloud services
with the AWS Management Console. With this web-based user interface
services can be created, modified and removed on demand by the customer.
These services are shared by multiple customers and the resources
provided can be scaled up and down. All statistics of used services
and resources and resulted cost are available directly to the customer~\cite{Mell:2011:SND:2206223}.

\subsection{Cloud Computing Elements}

The services in the cloud can be classified by service and deployment
model.

\begin{figure}[ht]
\centering
\includegraphics[width=\columnwidth]{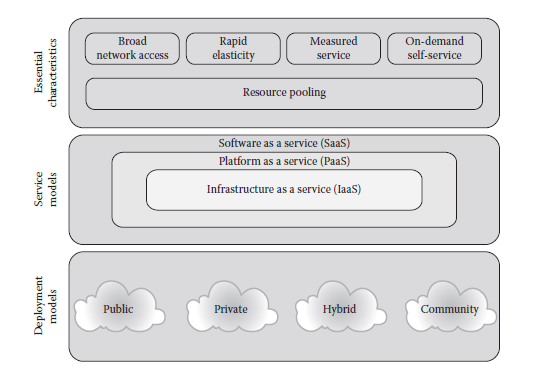}
\caption{Cloud computing elements~\cite[pp. 14]{Vacca:2016:CCS:3098688}}
\end{figure}

\subsubsection{Service Models}

As cloud service providers offer many different types of services,
these services are roughly categorized into three service models:
Infrastructure, platform and software. These models can also be seen
as layers, where basic services like providing storage, network or
computing services are on the bottom, supporting services in the middle
and specialized application services on top. The higher the service
in the layer, the more specific the service is, where the lower a
service, the more flexible usable it is. Normally services are using
other services of the same layer or layers below. A database service,
for example is using infrastructure services like storage and the
database software will normally run on an operating system in a virtual
machine. It will need additional services, for example access to a
domain name service, a firewall or the like.

\paragraph{Infrastructure as a Service (IaaS)}

Virtualized hardware is provided over a network. Storage or computing
resources are made available. Everything else, beginning from the
operating system is managed by the customer.

\paragraph*{Platform as a Service (PaaS)}

These services are provided for developers and administrators. They
can deploy applications on a software stack which is provided for
the customer. The service stack is managed by the cloud provider,
while the application itself is managed by the customer. Amazon (AWS)
also uses the terms ``Abstracted Services'' and ``Container Services''
for this service model.

\paragraph*{Software as Service (Saas) }

The applications of the provider are used by the customer. The model
is focused of the end user of the cloud. The user does not manage
or control the underlying cloud infrastructure, only user-specific
application settings are possible. Examples for this services are
Office 365 from Microsoft or Photoshop CC from Adobe.

\subsubsection{Deployment Models}

Another categorization is by deployment. This categorization is orthogonal
to the service model. Every service model can deployed in different
ways, and a deployment model does not have any influence on the service
model. The deployment has an direct impact on the security and privacy
of the data in the cloud, as there is an additional layer of defense
for the private and hybrid cloud, because the access to it is restricted.
Multiple service models can be deployed by using the same deployment
model, but the opposite is possible too, a service can be available
for multiple deployment models.

\paragraph{Public cloud}

The cloud infrastructure is provisioned for use by the general public.
This is considered the standard deployment scenario for cloud computing.

\paragraph{Private Cloud}

The cloud infrastructure is exclusively provisioned for a single (often
large) organization. This is the most secure deployment scenario,
as no unintentionally interactions with other customers or users of
the cloud are possible.

\paragraph{Community cloud}

The cloud infrastructure is exclusively provisioned for a specific
community.

\paragraph{Hybrid Cloud}

A composition of two deployment models. Although the cloud infrastructure
is separated, applications and data can be deployed in the private
and in the public part. A typical usage model is cloud bursting: If
the demand for resources exceeds the capacity of the private cloud,
these additional resources are provided by the public cloud. Azure
Stack makes it possible, to deliver cloud services from an internal
data center~\cite{Mell:2011:SND:2206223}.

\paragraph{Virtual Private Clouds (VPC)}

A private cloud is simulated in the public cloud, by physically isolating
storage and networking~\cite[pp. 43]{cloud}.

\subsubsection{Actors and roles in the cloud}

Outsourcing processes and services requires new actors and roles.
According to~\cite{cloud} the actors facility manager, service provider,
cloud user, and IT manager exist. Facility managers are the operators
for the outsourced services of the data center. These data centers
can be autonomous or a direct subdivision of the cloud vendor. The
service providers manage the resources of the data centers and are
employees of the cloud vendor. Cloud users are the customers of the
cloud services, they can be different from the end user. End users
are the people using the services provided by the cloud. IT Managers
are the people responsible of the computer infrastructure in an organization~\cite{cloud}.

\subsubsection{Cloud Computing Strength and Weakness}

Cloud computing is still growing fast. The main driver for this growth
are advantages over the classic in house operating and maintenance
model. According to~\cite{cloud},\cite{DBLP:journals/computers/KhalilKA14},\cite{ibm_cloud}
these advantages are:
\begin{itemize}
\item Reduced cost with shared infrastructure.
\item Avoiding over provisioning for peak times.
\item Eliminate the need to hire or train specialized IT staff for each
application and system.
\item Pricing models like charge per use and pay per use, resulting in more
budget flexibility.
\item On-demand elastically and scalability, resulting in more business
agility.
\item Any-where any-time accessibility.
\item Outsourcing of hardware and software management.
\item Better security.
\end{itemize}
Reduced cost is always a good motivation. But it is also good to know,
where this reduced cost come from. For cloud services the reasons
are economy of scale and shared infrastructure. Considering the size
of the big cloud vendors, economy of scale does not need any further
explanation but an even stronger effect comes from shared infrastructure.
Multiple customers can share the same infrastructure, but not only
infrastructure itself, but also the cost of maintenance for the infrastructure.
This minimizes capital investment in IT infrastructure and the need
to build out data center facilities. The provided pricing models are
very attractive too, as they are flexible and do not require any upfront
investments. This budget flexibility makes it possible to switch capital
expenditures for operating expenses. As the services are available
over the network, the cloud resources can be accessed from any location
and at any time. So a change of the region for a provided service
is simple and fast. Resizing the resources on-demand is a big advantage,
too, as peak demand for IT resources can easily satisfied. This makes
it possible to align costs with usage and to avoid over provisioning.
As result, this enhances business agility and makes it possible to
deploy and remove resources as needed. The next advantage can not
be overrated. Hardware and software management requires a lot of internal
resources in every organization. Any problems in this area directly
affects the performance of the whole organization. Security is often
a controversial advantage, because it is often overseen, that the
cloud has some benefits regarding security. First, due the availability
of more resources, denial of service attacks are much costlier for
the attacker, and chances are high that even DDoS (Distributed Denial of Service) attacks fail against
one of the major cloud providers. Second, another advantage is that
cloud provider usually have expert security personnel, who are specialized
for exactly the services they are running. This might not be affordable
in the in house data center. But of course, outsourcing into the cloud
has not only advantages. Some potential drawbacks and disadvantages
exist. These disadvantages are identified by~\cite{cloud},\cite{DBLP:journals/computers/KhalilKA14}
as:
\begin{itemize}
\item Loss of control on hardware and software.
\item Shared resource (performance reduced by neighbors).
\item Potential security risk by placing critical data on remote servers.
\item Vendor lock-in.
\end{itemize}
The loss of control can be seen as a disadvantage. The service provider
takes over control of hardware and software, whereas the degree of
control for the later depends on the chosen service model. The customer
therefore has no more control over the hardware and software used.
This means that any competitive advantages regarding software or hardware
are no longer possible. Shared resources can also have some severe
drawbacks. As already mentioned, sharing resources is great regarding
cost, but only as long as the use by other customers has no negative
impact on the provided resources. Privacy and security is a, if not
the, critical concern of outsourcing data. A lot of trust to the vendor
is necessary to give internal data out of hands. Depending on a specific
vendor for a service can also be a serious disadvantage, if the vendor
uses this dependency for its own benefit. Many of theses disadvantages
are in fact simply the other side of the coin of the advantages listed
before. It is great, not having to grapple with software and hardware,
on the other side it is not so great if you cannot control your software
and hardware anymore.

\subsection{Database as a Service }

\begin{figure}[ht]
\centering
\includegraphics[width=\columnwidth]{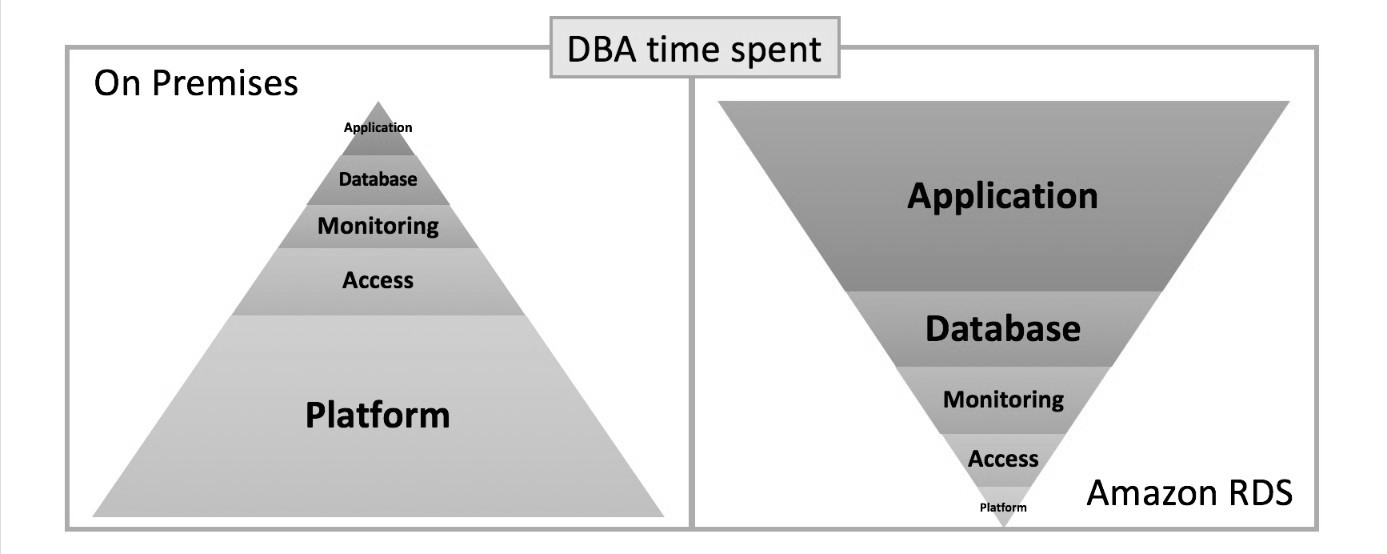}

\caption{\label{fig:Database-Administrator-Time}Database Administrator Time
(AWS)\cite{AWS-DBA}}

\end{figure}

As this report is about databases in the cloud, the database as a
service model (DBaaS) is examined in more detail. The database as
a service model was first described by~\cite{2002:PDS:876875.879015}
in 2002. It was implemented using the database DB2 and therefore named
NetDB2. Besides the performance overhead of remote access, it identified
data privacy as the most challenging problem and suggested first solutions
to the problem. A cloud deployment of a relational database has advantages
over the traditional in house approach. As already mentioned before,
cost is always a strong driver for migration into the cloud. In this
case it is not only the economy of scale on hardware and energy, but
also in workforce for administrating and maintaining the database.
\prettyref{fig:Database-Administrator-Time} shows the time spent
on different tasks by an administrator. Database administrators are
not waiting at every corner, and their expertise is well paid. Another,
often more important advantage is flexibility. Performance peaks are
easier to handle, because the cloud provider can easily distribute
the workload. Paying only the used resources is also a great benefit.
In short all general advantages of the cloud apply to the DBaaS model
as well. The commercial DBaaS offers have only the standard database
security features enabled. This means that normally the transport
is encrypted and the data is at least optional encrypted at rest.
To gain additional security, encryption on the application level is
necessary.

\subsubsection{Overview of DBaaS Providers}

All major cloud computing vendors offer managed relational databases
as a service. These services exist for both open and closed source
database management systems. The most widely used open source database
engines are MySQL/MariaDB and PostgreSQL. Multiple service provider
exist for the top commercial databases from Oracle and Microsoft ,
while a service using DB2 is offered only by IBM itself at the time
of writing. As there exist a myriad of constantly changing different
options of hardware and software options, here is only a short overview
of the provided services:

\paragraph*{Amazon Relational Database Service (RDS) }

Amazon offers the broadest range of services. Different options regarding
type and number of processors, amount of memory, size of storage and
network resources exist. (https://aws.amazon.com/rds/) The supported
database engines are listed below.
\begin{table}[H]
\caption{AWS DBaaS}
\resizebox{\columnwidth}{!}{
\begin{tabular}{|l|l|}
\hline
Database & Description\tabularnewline
\hline
\hline
Amazon Aurora & MySQL and PostgreSQL compatible\tabularnewline
\hline
PostgreSQL & PostgreSQL 9.3.12 - 10.5\tabularnewline
\hline
MySQL & MySQL 5.5, 5.6, 5.7\tabularnewline
\hline
MariaDB & MariaDB 10.0, 10.1, 10.2\tabularnewline
\hline
Oracle & Oracle 11g (11.2.0.4) and 12c (12.1.02)\tabularnewline
\hline
Microsoft SQLServer & SQL Server 2012 - 2017\tabularnewline
\hline
\end{tabular}}

\end{table}
%Amazon RDS with MySQL was used for the evaluation of Dice.

\paragraph*{Microsoft Azure Relational Databases}

Microsoft offers many different configurations for relational databases
in the cloud. These configurations vary regarding the used database
engine, the type and number of CPU cores and amount of memory provided
and the kind of deployment (shared, managed, single). (https://azure.microsoft.com/en-us/product-categories/databases/)
\begin{table}[H]
\caption{Azure DBaaS}
\centering
\resizebox{\columnwidth}{!}{
\begin{tabular}{|l|l|}
\hline
Database & Description\tabularnewline
\hline
\hline
Azure SQL Database & based on enterprise edition of SQL Server\tabularnewline
\hline
Azure Database for PostgreSQL & PostgreSQL 9.5, 9.6,10.4\tabularnewline
\hline
Azure Database for MySQL & MySQL 5.6, 5.7\tabularnewline
\hline
\end{tabular}}

\end{table}

Azure SQL Database was used during evaluation to show the support
of multiple databases (and not only MySQL).

\paragraph*{Google Cloud SQL}

Google offers a fully managed database service for (currently) two
open source database management systems. (https://cloud.google.com/sql/docs/)
\begin{table}[H]
\caption{Google DBaaS}
\centering
\begin{tabular}{|l|l|}
\hline
Database & Description\tabularnewline
\hline
\hline
Cloud SQL for MySQL & MySQL 5.6, 5.7\tabularnewline
\hline
Cloud SQL for PostgreSQL & PostgreSQL 9.6\tabularnewline
\hline
\end{tabular}

\end{table}

\paragraph*{IBM Cloud }

IBM offers its commercial databases DB2 and Informix as fully managed
service\footnote{https://www.ibm.com/cloud/databases}. Additionally it offers two open source relational databases
on its compose platform. Compose is a platform for all different kinds
of NoSQL and SQL databases and middleware like message brokers. This
includes services for MongoDB, Redis, JanusGraph, RabbitMQ , MySQL
and PostgreSQL. Apart from Db2 on Cloud, Db2 Hosted also exists as
the unmanaged version of DB2.
\begin{table}[H]

\caption{IBM DBaaS}
\resizebox{\columnwidth}{!}{
\begin{tabular}{|l|l|}
\hline
Database & Description\tabularnewline
\hline
\hline
Db2 on Cloud & a fully managed version of DB2 \tabularnewline
\hline
Informix on Cloud & https://www.ibm.com/cloud/informix\tabularnewline
\hline
Compose for MySQL & MySQL Version 5.7\tabularnewline
\hline
Databases for PostgreSQL & PostgreSQL Versions 9.4, 9.5, 9.6\tabularnewline
\hline
\end{tabular}}

\end{table}

\paragraph*{Oracle Database Cloud Service }

Oracle supports its own relational database in the versions 11g, 12c
and 18c as managed service\footnote{https://cloud.oracle.com/database}. Additionally Oracle offers its Exadata
Cloud Service, which is also available as customer edition. Exadata
is Oracles customized software and hardware for running its database.
The customer edition is the Exadata Cloud Service running at the customer's
own data center. MySQL is supported as cloud service, but not fully
managed.
\begin{table}[H]
\caption{Oracle DBaaS}
\resizebox{\columnwidth}{!}{
\begin{tabular}{|l|l|}
\hline
Database & Description\tabularnewline
\hline
\hline
Oracle DB Cloud Service & Version 11g, 12c and 18c\tabularnewline
\hline
Oracle DB Exadata Cloud Service & \multirow{1}{*}{Exadata }\tabularnewline
\hline
Oracle DB Exadata Cloud at Customer & Run at the customer's data center\tabularnewline
\hline
Oracle DB Exadata Express Service & Lightweight version of Exadata\tabularnewline
\hline
Oracle DB Schema Cloud Service & No full access to the database\tabularnewline
\hline
\end{tabular}}

\end{table}

%\newpage{}

\section{Security }

No software system with some useful functionality is without flaws
and bugs. Although these flaws and errors may restrict the use some
of its functionality, a system is often considered as working. If
the software is well designed, then an error in one component has
no impact on the usability of another component. The component which
is not affected, works as specified. In this aspect the security of
a system is different, because it is only as strong as its weakest
link. It does not matter, how secure one part of the system is, if
other parts, even only one, has a flaw, the whole security of the
system can be lost. Functionality can be tested by validating the
specification of its features. A system's security has to be tested
the other way, it is important to verify that certain functionality,
like accessing assets without authentication or authorization does
not exist in the system. This is significant harder to test than functional
features. Another difference is the adversarial setting. Normally
(legal issues set aside), a user wants to use the software in the
way it was intended by the developer, certain rules like constraints
and circumstances under which the software works are accepted. In
the context of security, there are no rules to which the attacker
has to comply. Another problem is that time is on the side of the
attacker. Attackers often can research and examine a system for years,
where the systems itself was developed under timing pressure and forced
to finish by an always too early deadline. All these things mentioned,
show that a secure system is hard to develop and even harder to maintain.
The more complex this software is, the harder it is to maintain security.
As mentioned in~\cite[pp 37]{Ferguson:2010:CED:1841202}, ``complexity
is the worst enemy of security''. Another important point to mention
is that security not only includes hardware and software, but also
its users and their interactions. Bruce Schneier brings this to the
point, saying ``Security is a process not a product''\cite[pp. XXII]{Schneier:2000:SLD:517959}.

\subsection{Privacy}

In nearly every country of the world, the gathering, using and transferring
of data, especially personal data is restricted by law. Every member
of the European Union has to have data protection laws, which implements
the directive 95/46EC of the European parliament and of the council.In
2018 this directive was replaced by the general data protection regulation
(GDPR) 2016/679 of the European Parliament and Council. The directive
restricts the processing of personal data in specific categories.
Examples of personal data are: Racial or ethnic origin, political
opinions, religious or philosophical beliefs, trade-union membership,
health or sex life.The processing of data in theses special categories
is generally forbidden and only allowed, if
\begin{itemize}
\item The subject has given explicit consent.
\item Processing is necessary for the purposes of carrying out the obligations
and specific rights of the controller in the field of employment law.
\item Processing is necessary to protect the vital interests of the subject,
where the subject is incapable of giving his consent.
\item Processing is carried out by an association with a political, philosophical
aim and that the data is only of members and not disclosed to others.
\item The processing relates to data which are made public by the data subject
or is necessary for legal claims.
\end{itemize}
In the United States of America the National Institute of Standards
and Technology (NIST) provides a guide to protecting the confidentiality
of personally identifiable information (Special Publication 800-122)
for all federal agencies. It defines personal identifiable information
as
\begin{quote}
``Any information about an individual maintained by an agency, including
(1) any information that can be used to distinguish or trace an individual\textquoteleft s
identity, such as name, social security number, date and place of
birth, mother\textquoteleft s maiden name, or bio-metric records;
and (2) any other information that is linked or linkable to an individual,
such as medical, educational, financial, and employment information.''\cite{mccallister2010sp}
\end{quote}
Examples of personal identifiable information are name, social security
number, driver license number, passport number, credit card number,
address information, images of a person, fingerprints or other bio-metric
data. The main recommendations for the handling of personal identifiable
data are to
\begin{itemize}
\item Minimize the use, collection and retention to what is strictly necessary
to accomplish their purpose.
\item To categorize them by the confidentiality impact level.
\item To apply the appropriate safeguards based on their confidentiality
impact level.
\item Develop an incident response plan to handle breaches.
\item Encourage close coordination between chief privacy officers, chief
information officers, chief information security officers and legal
counsel.
\end{itemize}
General speaking, the European legislative is more restrictive than
the American, because every member of the European union has to implement
the directives. The previous mentioned NIST guide for example is only
a recommendation and not a law~\cite{hanschdirective,mccallister2010sp,dsgvo}.

\paragraph{General Data Protection Regulation (GDPR)}

For cloud computing the responsibilities are split. The data processor
is the cloud computing service provider like Amazon (AWS) or Microsoft
(azure). The Data Controller is the customer of these cloud services~\cite{Voigt:2017:EGD:3152676}.

\paragraph{Information Systems Categorization regarding security}

Information and Information Systems can be categorized in three categories
regarding the objectives of security: Confidentiality, Integrity,
and Availability (also called the CIA triad)~\cite{Scarfone:2008:SGG:2206207}.
%~\cite{Processing_fipspub,Scarfone:2008:SGG:2206207}

\paragraph*{Low }

The loss of confidentiality, integrity and availability has only a
limited adverse effect, like minor damages or loss of assets on the
organization.

\paragraph*{Moderate }

The loss of confidentiality, integrity and availability results in
severe negative effects, like significant financial loss, but the
organization is still able to perform its primary functions.

\paragraph*{High }

The potential impact of the loss of confidentiality, integrity and
availability is high, if the organization is not longer able to fulfill
its primary functions or results in major damage and losses for the
organization or severe harm to individuals.

\paragraph*{Server Security Principles }

The NIST ``Guide to General Server Security''~\cite{Scarfone:2008:SGG:2206207}
also lists security principles like simplicity, fail-safe, complete
mediation, open design, separation of privilege, least privilege,
psychological acceptability, least common mechanism, defense in depth,
work factor and compromise recording. All these principles are applicable
to servers in the cloud, too. The difference is that some of these
principles have to be followed by the cloud vendor, and not the cloud
customer or user anymore.

\subsection{Cryptography}

\paragraph*{Cipher Security}

\begin{figure}[ht]
\centering
\includegraphics[width=\columnwidth]{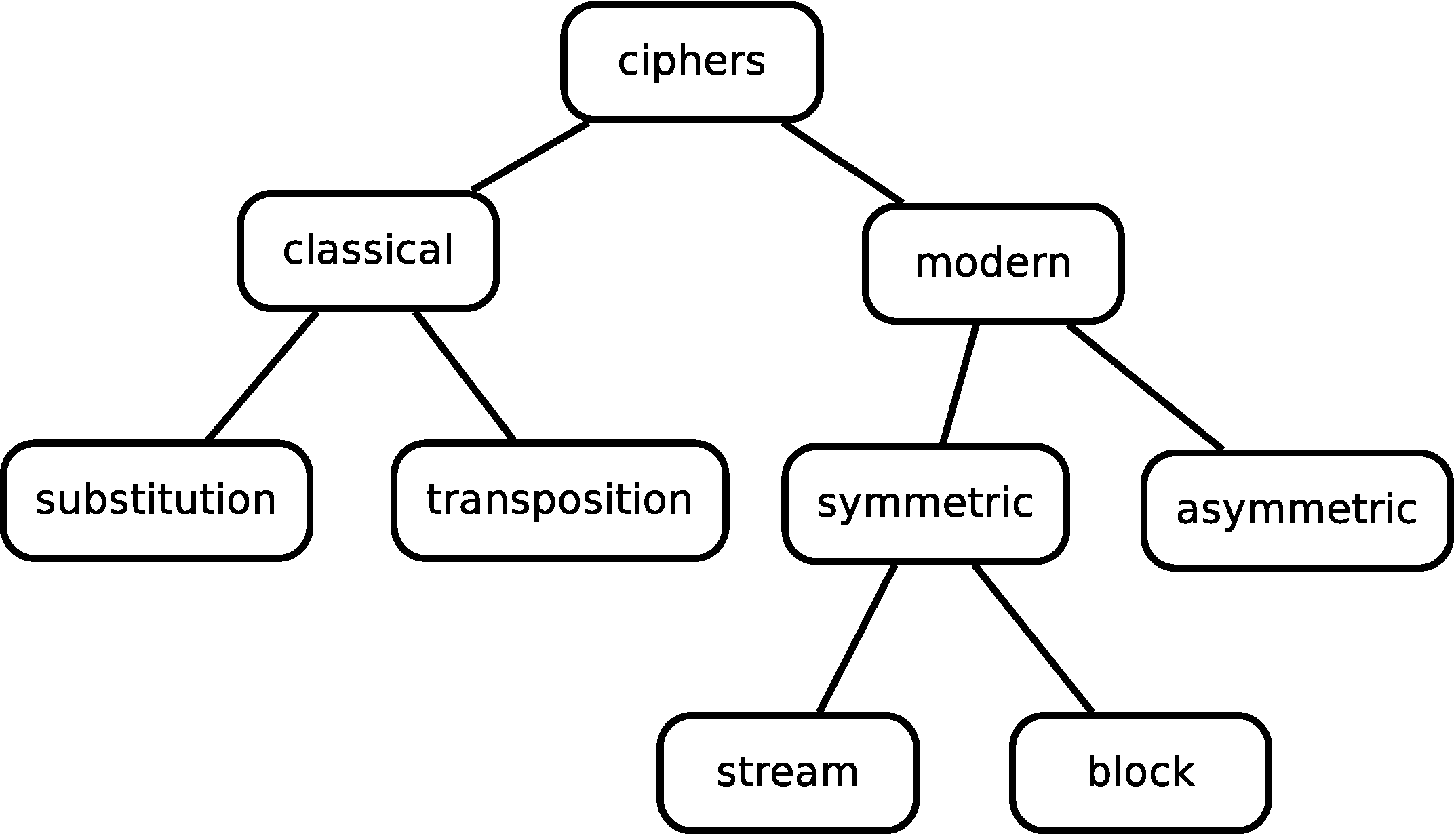}\caption{Taxonomy of ciphers}\label{fig:taxonomy}
\end{figure}

The security of a cipher is determined on how hard it is to break,
and how much information is needed to do so. An algorithm is unconditional
secure only, if no matter how much information the attacker has, it
is not possible to recover the plain text. If an algorithm is only
computationally secure, it means that it cannot broken by the available
resources~\cite{Schneier:1995:ACP:212584}. Cryptography is a vast
discipline with many ciphers. Many taxonomies of these ciphers exist,
but to give these ciphers some order and to give an overview a common
taxonomy was chosen (see \prettyref{fig:taxonomy}).

\subsubsection*{Kerckhoff's principle}
\begin{quote}
The security of the encryption scheme must only depend on the secrecy
of the key, and not on the secrecy of the algorithm\cite[pp 24]{Ferguson:2010:CED:1841202}.
\end{quote}
The opposite of this principle is called security through obscurity.
By security through obscurity an unknown cipher is used, or it is
even unknown that any cipher at all is used. The problem is, if the
obscure cipher is revealed, chances are high that it is not secure
anymore, because an detailed cryptoanalysis in public was never done.
Security through obscurity is not recommended anymore. NIST Guide
to General Server Security (SP 800-123) says regarding open design:
``Security should not depend on the secrecy of the implementation
or its components''~\cite[pp 2-4]{Scarfone:2008:SGG:2206207}. Steganography
in contrast to cryptography tries to hide secretly information in
plain sight. This can be also seen as security through obscurity.

\subsubsection*{Block ciphers}

Block ciphers are rarely used directly, because the message size usually
does not fit to the exact size of the block. For this are block cipher
modes, which support arbitrary message sizes. The simplest definition
of a secure block cipher according to~\cite[pp 44]{Ferguson:2010:CED:1841202}
is: ``A block cipher is secure, if it keeps the plain text secret.``
Another definition for security is the idea of an ideal block cipher.
An ideal block cipher means, that for each key value there is a random
permutation.
\begin{quote}
An ideal block cipher implements an independently chosen random even
permutation for each of the key values~\cite[pp 50]{Ferguson:2010:CED:1841202}.
\end{quote}

\paragraph{Block cipher modes}

To encrypt messages, that are not exactly one block long, a block
cipher mode has to be used. A block cipher mode is a function to encrypt
an arbitrary length plain text to an ciphertext. As often the size
have to be a multiple of the block length reversible padding is required
to get the original message after decryption. Some of the most common
modes for block ciphers are: ECB, CBC, Fixed IV, Counter IV, Random
IV, Nonce-Generated IV~\cite{Knudsen1998}.

\begin{itemize}

\item{ECB (electronic code book)}. Each block is encrypted separately in this mode. This means that two
identical blocks in plain text are encrypted to the same ciphertext,
which makes this mode less secure, because with a chosen plaintext
attack, it is easy to gain more information than acceptable. For example,
if records contain an encrypted name, and an attacker tries to reveal
if a distinct person with this name is in the record, than all the
attacker has to do, is to create a record with the same name, and
then select all rows with the encrypted name. Of course, sometimes
this deterministic property is exactly what is wanted. An example
is the foreign key relation in a database. If the same value is encrypted
to different ciphertexts, it is not possible to reference it as a
foreign key.

\item{CBC (cipher block chaining)}. Before encrypting the plain text, it is xor-ed with the previous encrypted
block in this mode. This avoids the problem of identical plain text
blocks resulting in the same ciphertext blocks. For the calculation
of the first XOR there is no previous ciphertext block, so an initializing
vector is needed. The following modes use different strategies to
solve this problem.

\item{Fixed IV}.
Here the initialization vector (IV) is fixed. It has the same disadvantage
as the ECB mode but for the first block only.

\item{Counter IV}. 
In this mode, the initialization vector is the sequence of the message.
This has the disadvantage that sometimes the sequence differs only
one bit which means, that the same plain text is encrypted in a very
similar ciphertext.

\item{Random IV}. 
Using a random initialization vector is secure, but has the disadvantage,
that this initialization vector has to be send to the receiver in
the first block. Therefore the ciphertext is one block longer than
the plain text.

\item{Nonce-Generated IV}. 
Instead of sending the initializing vector itself, only the information
(message counter) for creating an unique number (the nonce) is sent.
This normally requires less overhead than the IV itself.

\item{Stream cipher modes}.
Multiple modes like OFB, CTR, OCB, CCM, CWC and GCM exist for stream
ciphers.

\item{OFB (output feedback mode)}.
Here an output stream with the key is generated, and the plain text
is xor-ed with it.

\item{CTR (counter mode)}.
Another stream cipher mode, but in this case it is using a nonce.

\item{OCB,CCM,CWC and GCM}.
These are special modes combining encryption and authenticity functionality
at the same time.

\end{itemize}

\subsubsection{Classical ciphers}

All ciphers invented and used before 1950 are called classical ciphers.
They can be further classified in substitution and transposition ciphers~\cite[28]{ertel2007angewandte}

\paragraph{Substitution ciphers}

A substitution cipher is a cipher where every letter of the plain
text is substituted by a different letter. If every letter is always
mapped to the same encrypted letter, a substitution cipher is called
monoalphabetic.

\paragraph{Caesar cipher:}

The most famous and in fact one of the simplest of these ciphers is
the ``Caesar Cipher''. Each cleartext letter is replaced by shifting
the letter in the alphabet by n positions. The distance between the
letter and the substituted letter in the alphabet is always constant..
More general, the key in this substitution cipher is the number of
the shifts between the cleartext and ciphertext. Each letter of the
alphabet is assigned a number according to its position in the alphabet,
starting with letter A = 1 and ending with letter Z = 26. The function
for encryption can be written as:
\[
encrypt_{\text{K}}(P)=(P+K)mod|A\text{|}
\]
P is the plain text (respectively the position of a plain text letter
in the alphabet), K is the chosen key (number of shifts) an |A| is
the number of letters in the alphabet. To encrypt a message, for each
letter its position in the alphabet is determined, and the position
of the encrypted letter is calculated by adding the value of the key
modulo the number of letters in the alphabet. At the end the number
is replaced with the corresponding letter. For decryption the function
looks similar:
\[
decrypt_{\text{K}}(C)=(C-K)mod|A|
\]
In this function, C is the ciphertext and the other variables have
the same meaning as in the function for encryption. To decrypt the
message, the same steps are performed as in the encrypting function,
but instead of adding the value of the key, the key is subtracted.

Example: Given the key K = 3 (the original key used by Julius Caesar)
and the plain text ``VENI VIDI VICI'' the ciphertext can be easily
generated using the encrypt function: The first letter is 'V' so its
position in the alphabet is 22. Adding the key (3) results in 25 modulo
26, which is 25, respectively the letter 'Y' according to the chosen
alphabet. The next letter is 'E' encrypted to the letter 'H'. To make
it handier, a table of the alphabet in plain text and ciphertext can
be generated. The first row shows the positions of the letters in
the alphabet, the second shows the letters of the plain text and in
the third row the letters of the ciphertext can be seen.
\begin{table}[ht]
\caption{Caesar cipher with key = 3}
\resizebox{\columnwidth}{!}{
{\tiny{}}%
\begin{tabular}{|c|c|c|c|c|c|c|c|c|c|c|c|c|c|c|c|c|}
\hline
{\tiny{}position} & {\tiny{}1} & {\tiny{}2} & {\tiny{}3} & {\tiny{}4} & {\tiny{}5} & {\tiny{}...} & {\tiny{}9} & {\tiny{}10} & {\tiny{}11} & {\tiny{}12} & {\tiny{}13} & {\tiny{}14} & {\tiny{}...} & {\tiny{}22} & ... & {\tiny{}26}\tabularnewline
\hline
\hline
{\tiny{}plain text} & {\tiny{}A} & {\tiny{}B} & {\tiny{}C} & {\tiny{}D} & {\tiny{}E} & {\tiny{}...} & {\tiny{}I} & {\tiny{}J} & {\tiny{}K} & {\tiny{}L} & {\tiny{}M} & {\tiny{}N} & {\tiny{}...} & {\tiny{}V} & ... & {\tiny{}Z}\tabularnewline
\hline
{\tiny{}ciphertext} & {\tiny{}D} & {\tiny{}E} & {\tiny{}F} & {\tiny{}G} & {\tiny{}H} & {\tiny{}...} & {\tiny{}L} & {\tiny{}M} & {\tiny{}N} & {\tiny{}O} & {\tiny{}P} & {\tiny{}Q} & {\tiny{}...} & {\tiny{}Y} & ... & {\tiny{}C}\tabularnewline
\hline
\end{tabular}{\tiny{}
}}
\end{table}
To encrypt a letter, it is simply looked up in the plain text row
and substituted with the letter of the ciphertext. After the substitutions
'V' -> 'Y' , 'E' -> 'H', 'N' -> 'Q', 'I' -> 'L' the resulting ciphertext
of the first word in the example is then ``YHQL''. As letters which
are not members of the alphabet (in this case space) are ignored the
whole message encrypted is ``YHQLYLGOYLFL''. To decrypt it, each
letter is looked up in the ciphertext text row, and substituted with
the plain text, resulting in ``VENIVIDIVICI'', which is (after including
the corresponding spaces) the original message ``VENI VIDI VICI''.

A special case of this kind of cipher is ROT13. In this case the key
is 13 and the plain text is revealed if the cipher is applied on the
ciphertext a second time. The key space of these simple substitution
cipher is 26, which means that only 26 different keys for this algorithm
are possible. (Including the not very useful key 26, where
the ciphertext is equal the plain text.) The result is that the encryption
is very weak, and can be easily broken by an ciphertext-only attack.
Even if the alphabet would be larger (resulting in more possible keys),
it is easy to make a statistical attack on the ciphertext. When the
frequencies of the letters in a language are known, and the frequencies
of the encrypted letters are similar as in the plain text, it is easy
to guess which letter is mapped to the encrypted one. In the German
language for example, the letter ``E'' is with 17,4\% the most used
letter of the alphabet, so the probability is high, that in the ciphertext
the most frequent used letter is the encrypted ``E''. One solution
to avoid this kind of statistical attack is to map one plain text
number (the position of a letter) to one or more ciphertext numbers
according to their distribution in the language. These ciphers are
called homophone~\cite[29]{ertel2007angewandte,schmeh2009kryptografie}.

\paragraph{Multiplicative ciphers}

Another kind of monoalphabetic ciphers are multiplicative ciphers.
In a product cipher every letter (its position in the alphabet starting
with 0) is multiplied by a number, this number and the number of letters
in the alphabet has to be relatively prime, to make the decryption
unambiguous. The function to encryption and decryption can be written
as:
\[
encrypt_{\text{K}}(P)=(P*K)mod|A\text{|}
\]

\[
decrypt_{\text{K}}(C)=C*(Kmod|A|)^{-1}
\]
To decrypt the ciphertext has to divided modulo |A|, which is in fact
a multiplication with the multiplicative inverse modulo |A|, which
can be easily guessed for a small alphabet or calculated with extended
Euclidean algorithm. A weakness of multiplicative ciphers is the tine
key space. For an alphabet with 26 letters, there exist only 12 valid
keys, which vulnerable for a brute force attack~\cite{ertel2007angewandte}.

\paragraph{Polyalphabetic substitution ciphers}

A polyalphabetic cipher has not only one key, which maps a plain text
letter to the ciphertext but has many keys. For encryption of the
first letter the first key is used, for the second letter the second
key and so on. After the last key is used the circle starts again,
encrypting the next letter with the first key. The advantage of these
kind of ciphers is, that in the resulting ciphertext the distribution
of the letters is hidden, which complicates a statistical attack on
the cipher. Members of this kind of cipher are the Vigenere cipher
and the Hill cipher~\cite{Schneier:1995:ACP:212584,ertel2007angewandte}.

\paragraph{Transposition ciphers}

In a transposition cipher the difference between plain text and ciphertext
is only the order of the letters. A simple example of a transposition
cipher is to write down the plain text in rows of fixed length and
the ciphertext is the same text read by column. As there are many
other possibilities in which order the ciphertext can be read from
the table, many different ciphers of this kind exist. Another form
of a transposition cipher is called permutation cipher, where the
key is a permutation. All these ciphers are prone to statistical attacks,
because the distribution of the letters in the plain text is the same
as in the ciphertext~\cite{Schneier:1995:ACP:212584}.

\paragraph{Rotor machines}

These mechanical devices allowed the automatic encryption of a message
via a keyboard. It was a machine with a set of rotors in the end implementing
a polyalphabetic substitution cipher. The best known rotor machine
is the Enigma, which was used by Germany during WWII. As history showed,
the encryption was not bulletproof and the encryption was broken by
the British~\cite{Schneier:1995:ACP:212584}.

\paragraph{One-Time pad}

Although all of the classical ciphers are weak and can be easily broken,
there is one exception to this rule, the one-time pad. It is the only
cipher which is unconditional secure. The cipher is simple and depends
only on a good random key. This key consisting of random letters is
used to encrypt the message letter by letter. The letter from the
key is added to the plain text letter modulo 26. To decrypt the letter
of the key is subtracted from the ciphertext. The important thing
in this secure encryption scheme is, that every key is used only once.
Every message has to be encrypted with another key, and the key has
also be real random. The problem with this cipher is, that it is quite
unpractical, because the key has to be the same size as the message
(in fact, as the key can be only used once, a new key for every new
message is needed), both sender and receiver need the this key and
the key has to be truly random~\cite{ertel2007angewandte}.

\subsubsection{Modern Block ciphers}

\paragraph{Symmetric ciphers}

A symmetric cipher is a cipher where the same key is used for encryption
and decryption.

\paragraph*{Block ciphers}

Block ciphers are symmetric ciphers operating on a block of fixed
size. The plain text message is divided in blocks and each block es
encrypted separately.

\paragraph{DES}

Data Encryption Standard~\cite{Schneier:1995:ACP:212584}, also known as Data Encryption Algorithm
(DEA), was the first cryptographic algorithm, which became a ANSI standard.
DES is a block cipher with a block size of 64bit. The result of applying
the encryption on a block of plain text, is the ciphertext with the
same size. The length of its key is 64 bits but because every eighth
bit is used for parity checking the effective size is only 56. The
basic building blocks of DES are simple (which makes it easy to implement),
in fact only XOR, permutation and substitution is used. The application
of a substitution followed by a permutation is called round.

%\begin{figure}[ht]
%\centering
%\includegraphics[width=\columnwidth]{des}
%\caption{DES overview}
%\end{figure}

The algorithm works as follows: After an initial permutation (IP),
the block is split in two halves. Then 16 rounds (permutation and
substitution) are performed, in which the key is combined with the
data (in function F). At the end a final permutation is performed,
which is the inverse of the initial permutation.

The interesting part of the algorithm is the function F, where the
key (in fact a sub-key) is applied. In the first step of this function
a sub-key of length 48 is generated. This is called a compression
permutation, where the key is shuffled and reduced in one step. Then
the right half is expanded to length 48. This is done via a expansion
permutation, which not only changes the order of the right side, but
adds additional bits too. In the next step XOR is applied on the compressed
key and the expanded right side. The substitution is performed on
the result. DES has eight different substitution boxes (S-Box), each
having an input of 6 and an output of 4 bits. The 48 bit block is
distributed to the S-boxes and the substitution is performed. In the
next step, the 32 bit output is permuted by a P-box. At the end of
the round XOR is applied on the output of the right side and the left
side and the sides are switched for the next round. Decrypting works
the same way, the only difference is, that the order of the sub-keys
is inverted.

The security of DES depends on the length of the key and the implementation
of the S-boxes. The S-boxes, although not perfect showed only minor
flaws, which can be avoided. The problem is, that the key with 56
bits is to short, opening the door for brute force-attacks. Another
flaw is that if you choose 0 as key then all rounds use the same key
and as encryption and decryption are the same, except from the order
of rounds this distinguishes the algorithm from a ideal block cipher.
Another property of DES is, that if you encrypt the complement of
the plain text with the complement of the key, you get the complement
of the ciphertext. 3DES is attempt to enhance the security 3DES by
using three keys in sequence to encrypt a block. This solves the problem
of the small key size, but not the problem of the small block size.
Of course encryption/decryption takes three times as long as with
DES~\cite{schmeh2009kryptografie,ertel2007angewandte,Ferguson:2010:CED:1841202}.

\paragraph{AES\label{par:AES}}

Because DES with a key of 56 bit was no longer secure, a new standard,
the Advanced Encryption Standard (AES), was created. It is a cipher
with a block size of 128 bit. The key can have a length of 128, 192
or 256 bit. According to the length of its key, AES performs 10, 12
or 14 rounds. In every round except the last the following steps are
performed. Subbytes implements an S-box, doing substitutions. ShiftRow
does some permutations analog to a P-box. As all this operations are
done on a 4x4 matrix the next step MixColumns changes the order of
the columns. AddRoundKey, the last step adds a sub-key to the matrix.
In the last round the step MixColumns is replaced by AddRoundKey.
As of today no security flaws of AES are known, so the only possible
attack is a brute-force attack on the key, which is even with the
smallest length not possible with current available hardware. This
has not to be true for the future. In fact there are already theoretical
attacks on AES with 192bit and even 256bit key length~\cite{schmeh2009kryptografie,ertel2007angewandte,Ferguson:2010:CED:1841202}.
%\begin{figure}[ht]
%\centering
%\includegraphics[width=8cm]{aes\lyxdot jpg}\caption{Advanced Encryption Standard(AES)~\cite[pp. 54]{Ferguson:2010:CED:1841202}}
%\end{figure}

\paragraph{Stream ciphers}

Stream ciphers are symmetric ciphers, where each bit is encrypted
one by one. Stream ciphers are often faster than block ciphers, but
their security depends on the randomness of the used keys. A famous
example of a real secure stream cipher is the one-time pad, which
was already mentioned before in the group of the classical algorithms.

\paragraph{RC4}

Another well known and used represent of stream ciphers is RC4, named
after Ron Rivest. This cipher uses a 8x8 S-Box, with permutations
in the range from 0 to 255. The permutation depends on the key, which
has no fixed size. Then according to a simple algorithm, a random
byte K is generated and XOR is applied on K and the plain text. The
same thing is done to decrypt the ciphertext. The algorithm is about
ten times faster than DES. If the key is long and random enough, this
encryption is quite strong. If not, like in the case of WEP (Wired
Equivalent Privacy) it is not secure~\cite{schneier1996angewandte}.

\paragraph{Asymmetric ciphers}

An asymmetric cipher is a cipher where different keys are used for
encryption and decryption. They are also called public-key algorithms,
because at least one key is public available. It is crucial for an
asymmetric cipher, that the private key can not be deducted from the
public key. The plain text can be encrypted with the public key ,
and this plain text can only be decrypted with the private key. The
classic example is to send an encrypted email. To do this, the public
key of the receiver is used by the sender (the sender has to know
this key) to encrypt the message. This message is then sent to the
receiver, and can only decrypted by the private key of the receiver.

\paragraph{RSA}

The first commercial public-key algorithms was RSA, named after Rivest,
Shamir, and Adleman. The algorithm works as follows: To generate the
both keys (public and private), two large primes p and q are chosen
and multiplied, resulting in n. Then a encryption key e is chosen,
which has to be relative prime to (p-1)(q-1). Then the decryption
key d is calculated by

\[
ed\equiv1mod((p-1)(q-1))
\]
The public key consists of the numbers e and n, and the private key
is e. To encrypt a message it is split in blocks smaller than n. A
block is encrypted with

\[
m^{e}mod(n)
\]
and decrypted with
\[
c^{d}mod(n)
\]
The security of RSA depends on the fact, that factoring large numbers
is computational costly, because the factoring of n is required to
get the decryption key. The key length is obvious very important and
has normally a length from 1024 to 4096 bit.

\subsubsection{Tokenization }

``Tokenization is the process of randomly generating a substitute
value, or token, that is used in place of real data, where the token
is not computationally derived in any way, shape or form from the
original data value''\cite{sym_token}. While the plaintext and the
token is stored local, only the tokens are stored in the external
database. The tokens can, but do not have to, preserve the type and
format of the plain text data. The drawback of this approach is, that
the whole data has to be stored locally, which requires additional
resources and security measures. This can contradict the advantages
of moving the data in the cloud. If the tokens are ordered or searchable,
the same security drawbacks as of order preserving encryption exist.
Access from outside can be another issue, which can, from a security
standpoint, enlarge the attack surface of the solution significantly.
Although having the same goal as encryption, a difference to encryption
is, that tokenization is a non mathematical approach. As no sophisticated
processing is required, it is usually more performant than the encryption
process. Encryption, on the other hand requires no storage (except
for the key). While the security of ciphers is often analyzed thoroughly,
for tokenization this is often not the case, leading to security through
obscurity. As a standard for tokenization X9.119 exists. Visa has
defined best practices for tokenization in~\cite{visa}. Best practices
are for example: Segment the tokenization system from the rest of
the network, give only authenticated users access the system and monitor
it tightly. The tokens should be distinguishable from the plain text.The
token generation should use a strong cipher or a one-way reversible
function~\cite{McAfee,sym_token}.

\subsubsection{Cryptographic attacks}

The goal of cryptography is to keep messages secure. Cryptoanalysis
on the other hand has the goal to break the encrypted message and
reveal its plaintext. Cryptology is the combined study of cryptography
and cryptoanalysis. Practitioners of this discipline are called cryptologist
and have normally a strong mathematical background. As the terminology
of the domain cryptology is not unambiguous a short terminology for
this report is given. A plain text (also called cleartext) is a readable
message. Through the process of encryption the plain text message
is concealed. The resulting encrypted message is called ciphertext.
The reverse process, which restores the plain text from the encrypted
message, is called decryption. A cipher is a cryptographic algorithm
used to encrypt and decrypt a message. If the security of a cipher
is based on keeping the way the algorithm works secret, it is called
restricted. The sum of the cipher, plaintexts, ciphertexts and keys
is called a cryptosystem~\cite{Schneier:1995:ACP:212584}. Although
the focus of this report lies on cryptography, cryptoanalysis is also
needed as the complementary part of it, determining the quality of
a cipher and the usability for its applications. As the goal of the
cryptoanalysis is to recover the original plain text, there a different
type of attacks a cryptanalyst can perform.

Cryptographic attacks can be classified by the kind of access the
attacker has to a system. A type of attack is also called attack model~\cite[p.90]{Schneier:2000:SLD:517959}.
The more prerequisites and information for an attack is needed , the
harder it is for the attacker to be successful. Or seen from the perspective
of the cryptosystem: A cryptosystem is more secure, if it can withstand
an attack where the attacker has all the information he can gather,
than a cryptosystem, which can only withstand attacks where the attacker
has only limited information about the system. A threat model and
threat analysis is a prerequisite for every project. A threat itself
is defined as: ``An action by an adversary aimed at compromising
an asset''~\cite[pp 21]{Ferguson:2010:CED:1841202}. According to
\cite{Ferguson:2010:CED:1841202}, the following attack models exist:

\begin{itemize}

\item{Ciphertext Only Attack}.
The attacker has only the ciphertext of one or several messages. The
goal of the attacker is to recover the plain text of the messages,
or even to recover the key used for encryption. This is the hardest
way for an attacker to break an encryption system. Modern algorithms
normally withstand these attacks, because often the only way to decrypt
the message is a brute-force attack to guess the key.

\item{Known-Plaintext Attack}.
In this attack model the attacker knows the plain text and the associated
ciphertext for some messages. The goal of the attacker is again the
decryption of other messages or even getting the key to get access
to all messages. The access to plain text and ciphertext is quite
common, if you think about standard messages in a protocol or the
scenario, where the same message is sent encrypted to multiple receivers,
in which case multiple receiver know the plain text. Another scenario
is where the plain text is revealed after some time, because the secret
is revealed to the public anyway, like a quarterly report of a public
corporation.

\item{Chosen-Plaintext Attack}.
In this kind of attack, the attacker does not only know some given
plain text and its associated ciphertext like in the previous attack
model, but can also choose the plain text which is encrypted. The
attacker has access to the resulting ciphertext. This is often the
case when the attacker can specify input values, which are saved encrypted
by an application. A simple example is if you can create an user with
a password for an application and have access to the database, where
the password is stored, too.

\item{Chosen-Ciphertext Attack}.
The cryptanalyst can choose different plaintexts and ciphertexts and
has full access to the associated ciphertext and plaintexts. This
gives the attacker more freedom for the attack, making it more powerful
than the chosen-plaintext attack, although it is not that common as
the later. The goal of this attack is still to recover the key~\cite[p.32]{Schneier:1995:ACP:212584}.

\item{Distinguishing Attack}.
In the attacks before, the goal was to get access to the plain text
or the key. In this sort of attack the goal is to get additional information
about a new encrypted message after observing several messages before.
This means that although the new message can not be fully decrypted
at least some information is disclosed.

\item{Side-Channel attacks (information leakage)}.
In this kind of attack timing information or electricity consumption
can reveal information. Although the cipher itself is not compromised,
valuable information is leaked. An example could be that the validating
of a passwords takes more time if the first letters are valid. This
makes it much easier for the attacker to guess a valid password.

\item{Generic attack techniques}.
Another group of attacks is called generic attack. These attacks can
not be avoided, because they a system immanent. A example is DRM for
audio or video files. It is always possible to make an analog copy
of the audio or video file, when it is played. Other examples of generic
attacks, which can hardly be avoided are:
\begin{itemize}
\item Birthday Attack. This attack is named by the fact, that if there are
23 people in a room, the chance that 2 have the same birthday is more
than 50\%. In general this means that duplicate values and collisions
are not as uncommon as expected. This fact is important, where the
security is based on the fact that values are unique, for example
for transactions.
\item Meet-in-the-Middle Attack. This is another collision attack similar
to the birthday attack. For this attack, parts of the possible keys
are generated and stored in a table with a known plain text, known
from the protocol. If a match is found, some valuable information
(worst case a key) can be revealed.
\item Related-key attack. This attack is applicable if a relation between
different keys exist. If this relation can be revealed other attacks
can be launched.
\end{itemize}

\end{itemize}

\paragraph*{Attack tree}

\begin{figure}[ht]
\centering
\includegraphics[width=0.8\columnwidth]{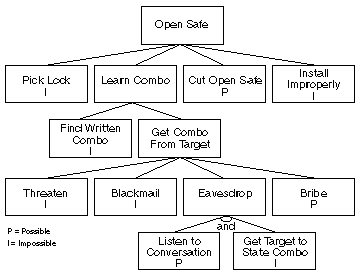}

\caption{\label{fig:Attack-tree-(Schneider)}Attack tree~\cite{schneier1996angewandte}}
\end{figure}

An attack tree can be built, to show the links of a security systems
and the possible attacks on them. \prettyref{fig:Attack-tree-(Schneider)}
shows the possible and impossible attacks on the different components
of a system from the perspective of the attacker. Other attributes
of the attack like cost and the needed know how can be added. The
goal is to think like the attacker and to fortify the links which
are protected the least.

\paragraph*{Forward secrecy}

Forward secrecy means, that even if a key is compromised, old ciphertext
is not revealed.

\subsection{Database Security }

In a database, the avoidance of unauthorized access to restricted
data is one of the key aspects of security. To prevent the disclosure
of sensitive data, a database management system provides a role-based
access control system to all objects in a database. With access control
it is possible to define and execute fine granular policies to ensure
the confidentiality of sensitive data. Although an important (maybe
the most important) single aspect in database security, it is not
enough to guarantee data confidentiality within the system. Defense
in depth as a security principle is always a good reason to provide
additional protection to data, as there are always chances that confidential
data can be leaked to unauthorized users. Threats like bugs in the
access control system or misconfiguration of user access rights can
never excluded completely. The even more important reason is, that
some threats against data confidentiality simply cannot be mitigated
by the access control of the database. This is specially true in the
case of deployment scenarios, where the provider of the database server
or the administrator of the database cannot or can only partially
be trusted. Common threats are:
\begin{itemize}
\item Eavesdropping of communication.
\item Legitimate privilege abuse.
\item Platform vulnerabilities.
\item Backup data exposure.
\end{itemize}
Data encryption can be a viable option to mitigate these threats.
Two main use cases for data encryption can be identified here: In
the first scenario the data is encrypted at rest and in the second
the it is encrypted in transit to avoid eavesdropping. Some solutions
provide data confidentiality in both scenarios, and mitigate multiple
threats, while some solutions only mitigate a single threat. In both
scenarios encryption plays a key role for data confidentiality~\cite{DBLP:reference/crypt/2011}.

\subsubsection{Encrypting data in transit}

Also known as encrypting data on the wire, encrypting data-in-transit
is mitigating the risk of eavesdropping the database communication.
Normally the communication between the database server and its client
is not encrypted and can easily be eavesdropped. To avoid this, the
communication between the server and the client or between database
servers for replication is encrypted. There is a number of ways to
achieve this goal by using a broad range of not only database specific
techniques. According to~\cite{Natan:2005:IDS:1211491} these techniques
can be summarized as:
\begin{itemize}
\item Database-specific features (Oracle Advances Security).
\item Connection-base methods (SSL).
\item Secure tunnels (SSH).
\item Relying on the operating system (IPSec).
\end{itemize}
The most common techniques used are SSL and IPSec and have a wide
range of applications, the encryption of database communication being
only one of them.

\paragraph{Secure Socket Layer}

SSL or more correct TLS (transport layer security) since Version 3,
is the de facto standard for e-commerce. It uses asymmetric encryption
for the exchange of a private symmetric session key, which is then
used for the rest of the session. Nearly all databases support the
use of SSL via the driver and an advantage is that TLS is protocol
independent and not restricted exclusively to the internet protocol,
because it is implemented above the transport layer.

\paragraph{IPSec }

IPSec (Internet Protocol Security) is another way to secure the communication
between client and database server. IPSec operates on a lower level
than SSL and can only used for IP traffic. IPSec is a standard and
supported by most modern operating systems. It can ensure the authentication,
integrity and confidentially of communication between two endpoints.
An advantage here is that it does not need the cooperation of the
applications involved in the communication, so a database server has
no support of IPSec, because it is not even aware that it is used
to encrypt its entire communication~\cite{IPSecHOWTO}.

\subsubsection{Encrypting data at rest}

\begin{figure}[ht]
\centering
\includegraphics[width=\columnwidth]{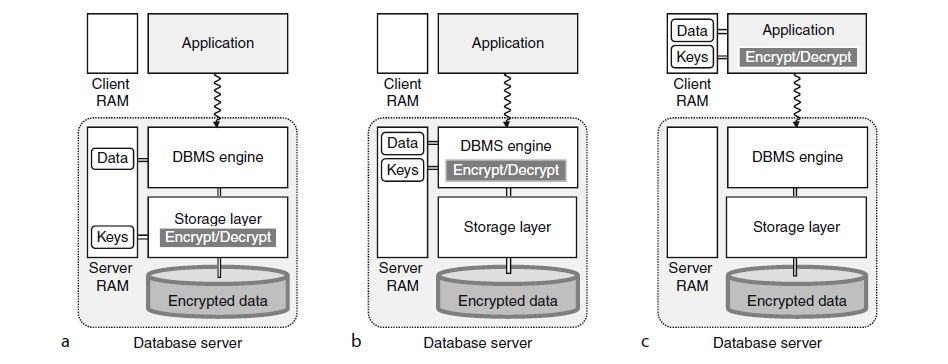}

\caption{\label{fig:db-level-encryption}DB-Level Encryption~\cite{Natan:2005:IDS:1211491} }

\end{figure}

Even if the database is perfectly secure and the database administrator
can be completely trusted, the data is still in danger, if the server
running the database itself is compromised. If the attacker gains
access to the file system by any means, it is possible to steal data
and log files of the database, and break its confidentiality. One
kind of threat is that the attacker gains physical access to the storage
system in cold state, either because of the rare case that the hardware
is stolen or the more common case that the data on old tapes or hard
disks is not completely destroyed. A solution against this threat
is to encrypt the file system of this disks, so even if physical access
can be gained, the confidentiality of the data is not broken. Another
threat exists, if the attacker gains access to the file system on
the running system. In this case the encryption of the file system
does not help, because during operation of the system, the files are
decrypted and can be accessed easily. As any vulnerability in the
software running on the server which enables access to the database
files is sufficient to attack, it is important to avoid the leakage
of confidential data even if access to the file system is gained.
This can be accomplished if the database encrypts all data, or at
least all confidential data in its files and log files. Most commercial
database vendors support the encryption of data and log files, and
they support the same symmetric ciphers for encryption. The biggest
differences in the solutions are the granularity of encryption, like
per column, table or database, and the management of the keys in the
database management system. The encryption is either built in the
database server or available as an option~\cite{Natan:2005:IDS:1211491,Cherry2012}. An
important decision is the place where the data at rest is encrypted.
It has an significant impact on the threats that can be mitigated
with the chosen solution. It also has great influence on the cost
and performance of the database. The levels where data can be encrypted
are:
\begin{itemize}
\item Storage-level encryption.
\item Database-level encryption.
\item Application-level encryption.
\end{itemize}
As shown in \prettyref{fig:db-level-encryption} the level, where
the encryption is performed has direct influence on the key management
and which threats can be mitigated on the specific level.

On storage level the file system is encrypted, minimizing the risk
that a backup discloses confidential information. Also the data and
logs are secure, if the server is in the cold state. If the server
is running, this is not the case, because any malicious application
on the server with enough privileges can access the data. The keys
are stored on the database server, normally managed by the operating
system. The database is not aware of the encryption. Other threats
are not mitigated.

When the data is encrypted at the database layer, the database transparently
encrypts and decrypts the data and logs. The keys are stored in the
database itself, which means that a database administrator has access
to keys and the encrypted data. During the operation of the database
management system the data is in clear text in the memory. As long
as the database application itself is not compromised, the data is
secured against backup disclosure and platform vulnerabilities. The
encryption and decryption puts a significant additional workload on
the database server. A big advantage is that the encryption is transparent
to any application, so there is no customization of the application
necessary.

If encryption is performed at the application layer, the database
server is completely excluded of the encryption process. The key management
is handled by the application and the encryption and decryption is
also performed by the application, meaning there is no performance
overhead for applying the ciphers on the server. As the size of encrypted
data is usually bigger than the clear text and some ciphers require
the selection and transfer of more data, this performance penalty
still applies on the server. All threats on the database server are
mitigated, but some like key management are only moved to the client.
A major disadvantage is that it is not easy to perform complex queries
with joins or aggregates on encrypted data on the database. There
are special ciphers with additional attributes which make complex
queries possible, but as already discussed, these solutions are either
not as secure or have an significant impact on performance and are
not as easy to use as standard symmetric block ciphers. An advantage
is that the application has the full control of the encryption and
can use in exactly the way needed.

\subsubsection*{Examples of encryption on database level }

\paragraph{MS SQL Server}

Microsoft SQL Server is supporting the encryption of all data and
log files of a database since Version 2008. This feature is called
``Transparent Data Encryption'' (TDE). TDE performs I/0 encryption
and decryption at the page level in real time. The data is encrypted
before written to the disk, and decrypted after it is read from the
disk. This is independent of the underlying file system. To do this,
the database encryption key is stored in the database and secured
by using a certificate stored in the master database. The protection
of this certificate and its private key is very important, because
the whole operation (including backups) of the database server depends
on it. The ciphers used for encryption are AES and 3DES with different
key lengths~\cite{TDE_SQLServer}.

\paragraph{Oracle Database}

Under the same name, but with different features, the Oracle Database
is supporting transparent data encryption since Version 10g, too.
It is part of an additional option called ``Advanced security option''.
This option includes the possibility to encrypt columns (constraint
by data types) and since Version 11g the encryption of whole table
spaces. For every encrypted table, a key is generated and all keys
are secured by a master key. The whole key management is done by the
database, and its master key is saved in a wallet or in a hardware
security module (HSM) for more security. The use of TDE on a table
or column is specified during the data definition in SQL. It is possible
to choose the cipher (AES or 3DES with different key lengths) and
the use of salt can be specified. Without the use of salt, every value
is encrypted to the same cipher text, so using salt is more secure,
but it has the limitation, that it is not possible to create an index
on encrypted column~\cite{TDE_Oracle}.

\paragraph{IBM DB2}

Although not called or advertised as transparent data encryption,
IBM has a solution for its database DB2 too. It is called IBM Database
Encryption Expert and can as TDE encrypt all data at rest. It supports
the encryption of backups and live data. Which data or files are encrypted
and which not, is configured by the encryption expert agent, which
can run on a separate server. As in the other solutions, the supported
ciphers are AES and 3DES,%~\cite{TDE_IBM}.

\paragraph{Vormetric Encryption for Databases}

This product is an example of a vendor independent solution, as it
supports Oracle, DB2, Informix, MS SQL Server, Sybase, MySQL and many
other database management systems. The encryption key management is
centralized and according to the vendor, it is completely transparent
to databases and applications. Add-on exist for encryption in the
cloud and for the Amazon cloud offering (AWS).

%\newpage{}

\subsection{Cloud Security }

\begin{figure}[ht]
\centering
\includegraphics[width=\columnwidth]{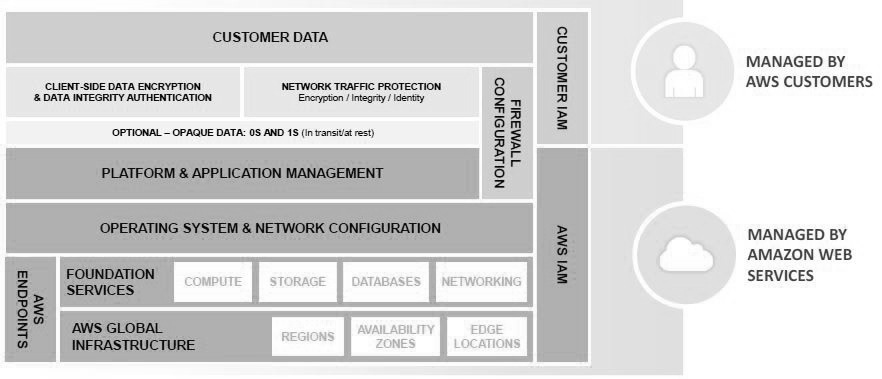}

\caption{AWS Shared Responsibility Model\label{fig:shared-responsibility-model}~\cite{aws_security} }
\end{figure}

According to~\cite[pp. 98]{cloud} and a recent survey from IDC
\cite{Nishad:2016:SPI:2905055.2905253} security is still the
biggest single concern in cloud computing. This is preventing or at
least slowing down cloud adoption. One of the problems in cloud computing
environments is the unclear definition of responsibilities between
the customer and cloud service provider. As stated in~\cite[pp 5]{Ferguson:2010:CED:1841202}
``A security system is only as strong as the weakest link'', it
is essential for security, that all different parts of a solution
are taken into account. For example a perfect maintained operating
systems does not help if the database has unpatched vulnerabilities
or is configured without any access restrictions. Cloud providers
like Amazon (AWS) try to solve this problem by defining the responsibilities
in a model they call ``Shared Responsibility Model''. This model
splits the responsibilities between the customer and the provider
by defining exactly for which configuration or artefact the customer
is responsible and for which the provider (AWS) itself. As these responsibilities
vary depending on the service model , each service type has its own
shared responsibility model. \prettyref{fig:shared-responsibility-model}
shows the shared responsibility model for container services. AWS
manages all underlying infrastructure up to the database. This includes
the operating system and the database management system itself. The
customer is responsible for the configuration of data backup / recovery
tools and the firewall. The responsibility for the customer data,
its encryption and import still lies in the hand of the customer~\cite[pp 9]{aws_security}.

Other important aspects of security in the cloud (and not only there)
are:
\begin{itemize}
\item Access Control.
\item Secure Communications.
\item Protection of private data.
\end{itemize}
Identity and authentication are fundamental for every secure software
system. This is especially true for services in the cloud, as all
resources are accessible over the network and due scalability there
are plenty of resources available. Authorization for access of each
asset is mandatory, too. As data has to be transferred in and out
of the cloud, secure transmission of data is essential for avoiding
eavesdropping of any kind. The protection of private data is fundamental
for every organization and required by law. To protect this data,
secure storage and computation is mandatory. Data integrity, confidentiality,
and the risk of data leakage are here the main concerns.

\paragraph*{Relational Database Security with AWS }

AWS supports some security features for its relational database service.
To get the data in and out of the cloud, encrypting the data in transit
is supported by most databases, as they support SSL/TLS wrappers.
This means that the driver communicates with the database in the cloud
via TLS. This feature is supported by all RDS MySQL and Microsoft
SQL instances. Amazon RDS for Oracle does not support a SSL/TLS wrapper
but has a native encryption solution called ``Native Network Encryption'',
which encrypts the data on transit. For the encryption of data at
rest the following options are available~\cite{aws_security}:
\begin{itemize}
\item Encryption on application level.
\item Oracle Transparent Data Encryption.
\item MySQL cryptographic functions.
\item Transact-SQL data protection function.
\end{itemize}
Encryption on application level does not need a support from the service
provider, as here the data is encrypted and decrypted by the application
and not the infrastructure service. Some databases like Oracle support
the transparent encryption of data at rest and some like MySQL and
Microsoft SQLServer have built-in functions to encrypt and decrypt
data.

\subsubsection{Security Concerns in the Cloud}

As already mentioned, security concerns prevent the adoption of cloud
solutions. According to~\cite{Chow:2009:CDC:1655008.1655020} these
concerns can be classified in:

\paragraph*{Traditional security concerns}

Computer and network intrusions are traditional security concerns,
which are relevant for the cloud, too. There aspects of the cloud
which can even reduce these concerns. For example cloud provider may
have better than average security measures and processes or insider
attacks are more unlikely, as security via contracts is easier than
internal control.
\begin{itemize}
\item VM-level attacks.
\item Cloud provider vulnerabilities (SQL-injections, cross-site scripting).
\item Phishing cloud provider.
\item Expanded network attack surface. Infrastructure to connect and interact
has to be secured.
\item Authentication and authorization has to be extended to the cloud.
\item Difficult forensics in the cloud, as there is limited access to equipment.
and media
\end{itemize}

\paragraph*{Availability}
\begin{itemize}
\item Uptime (arguments are they are as high as in house).
\item Single point of failure.
\item Assurance of computational integrity (is the application faithfully.
run and gives valid results)
\end{itemize}

\paragraph*{Third-party data control}
\begin{itemize}
\item Due diligence (Can cloud provider respond in the required time-frame
on a subpoena or other legal action, can deletion of data be guaranteed).
\item Audit-ability (Has the provider enough transparency for auditing purpose.
\item Contractual obligations by using the cloud.
\item Cloud provider espionage.
\item Data lock in (proprietary data format, vendor lock-in).
\item Transitive nature (the cloud provider has itself subcontractors).
\end{itemize}

\paragraph*{Information-centric security }

Focus is shifted from data from the outside (systems and applications
using the data) to protecting the data from inside. The data needs
to be self-describing and defending, regardless of the environment~\cite{Chow:2009:CDC:1655008.1655020}.
Data is encrypted and packaged with its access policy and when accessed,
reveals itself only in an trusted environment according to its policy.
Data owners wish to audit how their data is used and to ensure that
their data is not leaked or being abused. An approach is to have a
trusted monitor, that can provide proof on compliance to the data
owner. Another approach is to encrypt all data in the cloud, which
has the drawback of limiting data use. Searchable (predicate) encryption
is a method to restore the usability, even if the data is encrypted.
As of today no perfect encryption is available, but recent research
is promising. Other qualities like retrievability, which proves that
all data of a client is stored correctly, are important too. Cloud
fears largely stem from the loss of control of sensitive data. Control
can be extended to the cloud by using trusted computing and encrypting
all sensitive data~\cite{Chow:2009:CDC:1655008.1655020}. This is exactly
the attempt of this report, encrypting all sensitive data, while still
being usable in a database.

Providing access to a database in the cloud has, apart from the usual
requirements for security and privacy, additional needs, because the
data is not longer in the scope of the organization owning the data.
One of these requirements is for example the secure transport from
and into the cloud. In the cloud itself, the data has to be secured
too, but not only against attacks from outsiders, but additional from
insiders or the service provider too. Often this threat is not even
a malicious cloud provider, but curiosity alone is enough motivation
to breach privacy. As a public cloud normally has multiple clients,
strict isolation between their data is an additional mandatory requirement.

\subsubsection{Privacy and Security}

As the transfer from and to the cloud can be secured by standard technologies
like SSL/TLS or the like, the main goal is to secure the data, when
it is in the cloud. The best way to achieve this, is to encrypt all
or at least all sensible data in the cloud. For this, all techniques
described in the chapter database encryption are relevant, but there
are some additional things to consider, like compatibility with features
of the cloud approach, for example scalability. The later in this
chapter described project Relational Cloud is an approach to fulfill
all these requirements.

\paragraph{Attack Models}

According to the Cloud Security Alliance (CSA) the ``Treacherous
Twelve Cloud Computing Top Threats''~\cite{CSA} are:
\begin{enumerate}
\item Data Breaches. A data breach in general is an incident where confidential
data is accessed without authorization. The impact of the data breach
depends on the confidentiality and the scope of the disclosure of
the information.
\item Insufficient Identity, Credential and Access Management. If attackers
gain control of credentials and private keys security is broken. Therefore
credentials and cryptographic keys have to be secured in particular.
This includes activities to prevent weak passwords, invalid certificates
and lack of cryptographic key rotation.
\item Insecure Interfaces and APIs. The security of cloud services depends
on the provided application programming interface. This interfaces
must provide protection against circumvention of authentication and
access control.
\item System and Application Vulnerabilities. Every system and application
can have bugs and flaws which can be used to compromise the system
or application. Systems and application have to be updated and patched
against found vulnerabilities to prevent the risk of data breaches.
\item Account Hijacking, Accounts can be hijacked by attack methods like
phishing or simply social engineering. A compromised account can be
used as a base for another attacks. Make matters worse, credentials
are often reused, compromising not only one but multiple accounts.
\item Malicious Insiders. A malicious insider can be a great threat. This
is particular true in the cloud, because the insider of the service
provider is normally no member of the organization, which owns the
data.
\item Advanced Persistent Threats. A system is compromised for a long period
by using sophisticated techniques to exploit vulnerabilities. The
attacker in such cases can be a big organization or even a country
with much more power at hand, than a normal attacker.
\item Data Loss. Losing important and not restorable data is a big concern
in the cloud. Reasons for data loss are not only attacks, but also
natural disasters of all kind.
\item Insufficient Due Diligence. As the cloud service provider is normally
not part of the organization, which uses its services, all technical,
commercial, legal and compliance issues have to be evaluated, before
an outsourcing is possible.
\item Abuse and Nefarious Use of Cloud Services. If cloud services are not
secured, they can be abused to cause damage on other systems. Misuse
can be a DDoS attack, sending of spam mails or hosting illegal content.
\item Denial of Service. A denial of service attack prevents users from
using a service. As a service in a public cloud has to be available
on the internet and not only in an internal network, the attack surface
for this kind of threat becomes larger in the cloud.
\item Shared Technology Issues. A cloud service provider usually shares
its hardware and software for different customers. If these shared
resources are not completed isolated, than the confidentiality of
the data is at risk. In such a scenario it is possible that customer
A using the same cloud as customer B, gains access to information
of customer B.
\end{enumerate}
A risk is composed of a threat, a probability and an impact. According
to~\cite{assess_database} a taxonomy of risks is given:

\paragraph*{A Taxonomy of risks:}
\begin{itemize}
\item Organizational Risks like Loss of reputation, loss of share value.
\item Technological Risks Risks associated with the use of technological
services like design,engineering, processes and procedures.
\item Legal risks All issues due to legislation and regulations.
\item Human errors an accidents.
\item Network threats.
\item Application threats.
\item Host Threats.
\end{itemize}

\paragraph*{Threat assessments:}
\begin{itemize}
\item Data Threats (loss of data or integrity).
Attacks cracking authentication
credentials, SQL injection, privilege escalation, unpatched database
vulnerabilities, human errors, loss of encryption keys) These threats
are not cloud specific, but using the internet as medium widens the
risk (bigger attack surface) Another impact on security is less data
control in the cloud. Establishing auditing controls are more difficult
in the cloud. Data lock-in can also be seen as security issue. It
is important to know, what happens with the data if the provider is
changed.
\item Physical Threats. 
When natural disasters occur, infrastructure can
be harmed. As cloud providers often operate on larger scale, it is
possible that they have more barriers against physical threats (for
example datacenters in multiple regions). This threat is not cloud
specific, too.
\item Interface Threats. 
Are not only cloud specific, too. Security policy
should ensure authentication and access controls. Monitoring is more
complex in the cloud, as there are various layers of applications.
\item Authentication Threats (Phishing, etc). 
They are the same in the cloud,
but the impact in the cloud can be much higher, as there are often
multiple clients affected.
\item Virtualization Threats int the Cloud. 
Nearly all computing in the
cloud depends on the virtualization paradigm. Isolation is not guaranteed.
Attacks on hypervisors are possible.
\item Cloud Power Threats. 
Denial of Service can affect multiple clients
and applications can not separated from the internet.
\item Outage. 
Although the chance is less probably, the impact of an outage
in the cloud is much higher.
\end{itemize}
Traditional IT is easier to control and to manage, but also much more
limited in resources (technical and human). There is no right answer
whether traditional computing or cloud computing is better in terms
of security~\cite{assess_database}.

\paragraph*{Threat classification}

STRIDE is a threat classification model for computer security threats.
It is a mnemonic for
\begin{itemize}
\item Spoofing of user identity.
\item Tampering.
\item Repudiation.
\item Information disclosure.
\item Denial of service.
\item Elevation of privilege.
\end{itemize}
It is used to reason about threats and find threats to a system~\cite{jouini2014classification}.

\paragraph*{Threat Risk Modeling}

To assess the risk of a security threat and damage potential of it,
the DREAD model can be used. It stands for
\begin{itemize}
\item Damage (How much damage will be caused, if an exploit is successful?).
\item Reproducibility(How much effort is needed, to reproduce the exploit?).
\item Exploitability (What is needed to exploit the threat?).
\item Affected users (How many users a affected by the threat?).
\item Discoverability (How easy is it, to discover the threat?).
\end{itemize}
The last item of the list (discoverability) is controversial discussed,
because it can be interpreted as ``security by obscurity'', which
itself is controversial, and some say is no security at all, where
others say, that every obstacle for an attacker, and not knowing that
there is a threat is an obstacle, enhances security.
\begin{quote}
$Risk_{DREAD}=$(DAMAGE + REPRODUCIBILITY + EXPLOITABILITY + AFFECTED
USERS {*} DISCOVERABILITY) / 5
\end{quote}
Values from 0 to 10 are assigned to each attribute. 0 means nothing
or minimum effects and 10 maximum effects. The simple formula above
calculates the overall risk of a treat, the higher the score, the
higher the risk~\cite{owasp_dread}.

\paragraph*{Insider threats}

Many mechanisms exist for protecting data from outside attacks. Unfortunately,
these mechanisms fail to protect data from authorized users from inside,
who abuse their privileges. The protection of sensitive data from
insiders is as important as protection from outside, as the the adverse
consequences are the same. Insiders can use tables and relations they
are authorized to access. Furthermore they also may gain additional
knowledge by using dependencies. These dependencies can be:
\begin{itemize}
\item functional dependency.
\item fuzzy dependency.
\item multi valued dependence.
\end{itemize}
A functional dependency can reveal full information about another
field, which depends on an accessible field. Fuzzy dependencies only
reveals partial information, where as multi valued dependencies do
not leak information~\cite{10.1109/CSE.2009.159}. Replication and load
balancing of a database in the cloud can increase the probability
of insider threats, which cannot be easily detected. To mitigate the
threat the activities of insiders have to be monitored on different
instances in different zones. Knowledge bases of insiders have to
be monitored and synchronized. As this is normally the responsibility
of the data owner, in the cloud it is the responsibility of the
cloud provider. Different prevention models exist~\cite{10.1109/UCC.2012.18}:
\begin{itemize}
\item Peer-to-Peer.
\item Centralized.
\item Mobile Knowledge bases.
\end{itemize}

\paragraph*{Threat mitigation}

There are many possibilities to mitigate these threats like
\begin{itemize}
\item Data Obfuscation (Masking, Scrambling). 
Fake or scrambled data set
for use by design and implementation teams, can be very expensive.
\item Encryption of Data. 
Allows personally identifiable data to be scrambled
if intrusion takes place, but adds overhead and possible performance
issues.
\item Database Intrusion/Extrusion Prevention. 
Looks for SQL injections, bad access commands and odd outbound data. Can cause performance issues,
and needs very specific criteria to set up.
\item Data Leak Prevention. 
Catches any data that is being sent out of the
system. Does not protect data in the actual data warehouse.
\end{itemize}
Other threats are:
\begin{itemize}
\item Lost or stolen media. 
Can be mitigated by encryption.
\item Unauthorized file sharing. 
Can be mitigated by encryption.
\item Privileged user abuse. 
Can be mitigated by encryption, separation of duties and application authentication audit.
\item Data leakage/unauthorized access. 
Can be mitigated by policy-based security.
\end{itemize}
%\newpage{}

\section{Database Encryption in the Cloud}

As shown before, deploying a database into the cloud has advantages,
but privacy concerns remain. To solve this problem, it would be great
to only deploy the data encrypted and never as plaintext into the
cloud. So in short, the goal is to deploy a database in the cloud,
with all sensitive data encrypted, but still being usable as if the
data were not encrypted at all. No change of an application or a query
should be necessary.

\subsection{Prerequisites}

It is important to classify the data according to its sensitivity,
because sensitive data needs special protection. A naive solution
would be to classify all data as sensitive, but as this comes with
an overhead, it is often no viable solution. For handling sensitive
data different scenarios exist:

\subsubsection{No Sensitive data at all}

Maybe this looks like a trivial scenario, but often applications store
and process more data than they need. Nowadays the technical tools
(hardware and software) exist to process more data than ever, the
credo here is often: the more data the better. But this does not come
for free, and has some significant drawbacks regarding cost for storing,
retrieving and managing the data. If this data contains sensitive
data, then this cost is even much larger, because this sensitive data
has to be protected, and the protection mechanism often include big
overhead in the processing and management of the data. Also the legislative
environment is changing to more restrictive privacy protection laws,
and user and customer awareness of privacy issues is rising. So if
it is possible to implement an application without the need of the
processing and storing of sensitive data, this is simply the best
case.

\subsubsection{No Sensitive data in the cloud}

If sensitive data is needed another way to avoid the drawbacks is
to simple split the application and the database in two separated
systems. One system has no sensitive data at all and can be easily
deployed in the cloud like in the first scenario. For the system with
sensitive data, there are either one of the following encryption strategies
possible or the system can hold its data off the cloud. Of course
it is important that this system does not only store the sensitive
data separately, but also does not leak the information via its provided
services. If you have an architecture based for example on microservices,
then this can be a viable approach. Only services with no sensitive
data are deployed in the cloud. The approach of this scenario is quite
similar to the first, by rather not solving but avoiding the problem.

\subsubsection{Encrypting sensitive data in the cloud}

This is the case where at least some of the data is classified as
sensitive and is deployed encrypted in the cloud. The data in a database
can be encrypted on different levels of granularity. These levels
of granularity are~\cite{foresti2010preserving}:
\begin{itemize}
\item Relation.
\item Attribute.
\item Tuple.
\item Element.
\end{itemize}
If the relation is encrypted, all data of an table is stored in a
single value in the encrypted database. The granularity ``Attribute''
means, that each column is stored in a single value in the encrypted
database. If the encryption is tuple based, a whole row (or at least
most of it) is encrypted as one value in the encrypted database. The
finest granularity is where each element is encrypted as single field
in the database. As encryption of the levels relation and attribute
make any normal usage of the database impossible, encryption based
on tuple and specially on element granularity is preferred.

\subsection{Encryption Strategies}

To encrypt the data, multiple strategies are possible. Which to use
best is highly dependent on the data model, data usage (like updates
and kind of queries) and confidentiality or privacy requirements of
the data.

\subsubsection{Standard ciphers}

The sensitive data is stored encrypted with a standard symmetric block
cipher like AES. This is done without considering the datatype used
in the database scheme. The data or better queries on it cannot processed
in the cloud. Only exclusion is that the encrypted data can be selected
like any binary field in the database. To process the data in a meaningful
way, it has to be downloaded and decrypted. For range query this means
that all data has to be downloaded. This is only practical if there
are no such queries at all or these queries are executed only on a
very small amount of data, as in all other scenarios the performance
impact would be prohibitive. The key management is also a critical
part of this solution, because it has to be implemented on every private
server where the data is processed. The advantage of this scenario
is, that it is possible to use proven existing ciphers and can still
achieve some of the benefits of the cloud like availability and scalability.
As there is no processing possible in the cloud, it is more like a
central master database in the cloud, from which every client can
download or sync the data before it can be processed local.

\subsubsection{Standard ciphers and augmented data model}

In this scenario the data is fully encrypted as in the scenario before,
but for each encrypted data field additional information on the data
is provided. This can be additional information on order or equality
of a field. The data model is extended for these fields. This makes
range queries possible, by operating on the additional fields, and
not on the encrypted data itself. Of course it is important, that
this additional data does not leak information on the sensitive data
itself. If these additional fields have to be computed for every update
or delete, it could make the performance penalty prohibitive. For
example, if any addition of one row, requires the update of all existing
rows, to calculate the new orders in a row this would not work very
well. A major disadvantage is, that the data model differs significant
from the normal, domain specific data model and every query has to
use the additional fields to work on the data. The advantage on the
other side is that standard encryption technology can be used and
some of the restriction on queries can be avoided. As queries are
executed on a different data model, they have to rewritten to be compatible
for the enhanced data model. It is of course useful to automate the
rewrite of the query, either by generating these queries or by rewriting
it in a proxy.

\subsubsection{Ciphers with additional properties}

Beside from standard ciphers, other ciphers with additional properties
exist. Encrypting the data with these ciphers makes it possible, to
use the encrypted data as if it were plaintext. This strategy needs
no change in the data model, which sounds only too well. The drawback
is that there is a overhead of these ciphers and they are not as secure
(at least not nowadays) as standard ciphers. Which properties are
required is highly dependent of the queries used. Ciphers with properties
are described later in detail \ref{subsec:OPE-(order-preserving}
and \ref{sec:Property-Preserving-Encryption}, but before describing
them, it is essential to determine the requirements of queries on
the encrypted data.

\subsection{\label{subsec:Query-Requirements}Query Requirements}

As mentioned before, to perform non trivial SQL queries on encrypted
data, the used ciphers have to provide additional properties to work.
To make queries work on encrypted data without modification, different
properties have to be supported by the used encryption scheme. The
listed elements of queries are based on the description of relational
operators in~\cite{Date:2011:SRT:2208088}. But before examining the
queries, a short overview of the requirements originating from the
used datatypes is given.

\subsection*{Datatypes}

As SQL uses different types, it is necessary, that the encrypted data
is still a valid attribute. Format preserving means in this context
not a special domain format like a social security number, but more
a kind of type preserving encryption.

\subsubsection*{String datatypes (CHAR, VARCHAR, TEXT ...) }

If the field is long enough, no additional properties are required
for the cipher. If the length is too short or a special encoding is
necessary, format preserving encryption is required.

\subsubsection*{Number datatypes (TINYINT, INT, FLOAT, DECIMAL ...) }

Format preserving encryption is required.

\subsubsection*{Date datatypes (DATE, DATETIME, TIMESTAMP, TIME...)}

Format preserving encryption is required.

\subsection*{Restrictions}

A \texttt{select} is restricted by one or multiple conditions. Multiple operators
are supported, and the conditions can be combined with ``and'' and
``or``. These operators determine the properties required for a
cipher, so the are examined separately.

\begin{lstlisting}
SELECT T.* FROM Table T WHERE T.Attr OP Value
\end{lstlisting}
 or

\begin{lstlisting}
SELECT T.* FROM Table T WHERE T.Attr BETWEEN Value1 AND Value2
\end{lstlisting}

\begin{itemize}

\item{Equal (=)}.
Equal requires a deterministic cipher.

\item{Not Equal (``<>'')}.
Not equal requires a deterministic cipher.

\item{Less (``<'')}.
Requires an order preserving cipher.

\item{Less Equal (``<='')}.
Requires an deterministic order preserving cipher.

\item{Greater ('' >'') }.
Requires an order preserving cipher.

\item{Greater Equal ('' >='')}.
Requires an deterministic order preserving cipher.

\item{Between}.
Requires deterministic order preserving encryption

\end{itemize}

\subsection*{Projection}

\begin{lstlisting}
SELECT DISTINCT T.Attr1, T.Attr2 FROM Table T
\end{lstlisting}
For the elimination of duplicates (DISTINCT) a deterministic cipher
is needed. If no duplicates are eliminated, there are no requirements
for a cipher.

\subsection*{Join}

\begin{lstlisting}
SELECT T1.Attr1, T2.Attr2 FROM Table1 T1 JOIN Table2 T2 ON T1.Attr2 = T2.Attr2
\end{lstlisting}
As the join uses the equal operator a cipher has to be deterministic.

\subsection*{Union}

\begin{lstlisting}
SELECT T1.Attr1 FROM Table1 T2 UNION Select T2.Attr1 FROM Table2 T2
\end{lstlisting}
As union has an implicit distinct, filtering duplicate entries, a
deterministic cipher is needed.

\subsection*{Union All}

\begin{lstlisting}
SELECT T1.Attr1 FROM Table1 T2 UNION ALL Select T2.Attr1 FROM Table2 T2
\end{lstlisting}
Union all has no additional requirements for a cipher.

\subsection*{Intersect}

\begin{lstlisting}
SELECT T1.Attr1 FROM Table1 T2 INTERSECT Select T2.Attr1 FROM Table2 T2
\end{lstlisting}
For intersection a deterministic cipher is needed.

\subsection*{Difference}

\begin{lstlisting}
SELECT T1.Attr1 FROM Table1 T2 EXCEPT CORRESPONDING Select T2.Attr1 FROM Table2 T2
\end{lstlisting}
For difference to work, a deterministic cipher is needed.

\subsection*{Semijoin}

\begin{lstlisting}
SELECT T1.* FROM Table1 T1 WHERE T1.Attr1 IN (SELECT T2.Attr1 FROM Table2)
\end{lstlisting}
As the semijoin uses the equal operator a cipher has to be deterministic.

\subsection*{Semidifference}

\begin{lstlisting}
SELECT T1.* FROM Table1 T1 WHERE T1.Attr1 NOT IN (SELECT T2.Attr1 FROM Table2)
\end{lstlisting}
As the semidifference uses the equal operator a cipher has to be deterministic.

\subsection*{Extend}

\begin{lstlisting}
SELECT T1.Attr1 * 123, T1.Attr2 FROM Table1 T1
\end{lstlisting}
As an arithmetic calculation (+,-,{*},/,\%) is performed, the cipher
has to be homomorphic supporting the arithmetic operation.

\subsection*{Aggregate Operators}

\begin{lstlisting}
SELECT SUM(T1.Attr1) FROM Table1 T1 GROUP BY T1.Attr2 HAVING SUM (T1.Attr1 < 2000)
\end{lstlisting}
The cipher has to be homomorphic (addition) and order preserving for
the having clause, if another comparison operator like equal is used,
than the cipher has to be deterministic.

\begin{lstlisting}
SELECT MIN (T1.Attr1) FROM Table1 T1
\end{lstlisting}
The cipher has to be order preserving.

\subsubsection*{Group By }

The cipher for the relevant attribute has to be deterministic.

\subsubsection*{Having}

The same properties as for aggregate operator are needed.

\subsection*{Summarization}

\begin{lstlisting}
SELECT COUNT(*) from Table1
\end{lstlisting}
No additional properties are required.

\subsection*{Order By}

\begin{lstlisting}
SELECT T1.* from Table1 T1 ORDER BY T1.Attr1 DESC
\end{lstlisting}
The cipher has to be order preserving.

\subsection*{Like}

\begin{lstlisting}
SELECT T1.* from Table1 T1 WHERE T1.Attr1 LIKE '%String%'
\end{lstlisting}
The cipher has to be searchable.

\subsection*{Functions (Substr, Concat ...)}

The cipher needs to be homomorphic, and the Functions must be implemented
as user defined functions.

\begin{lstlisting}
SELECT SUBSTR(T1.Attr1,1,5) from Table T1
\end{lstlisting}

\subsection*{Summary}

To give an overlook of the properties the relational operators are
listed in \prettyref{tab:Required-Properties-for}. Of course depending
on the type of the used attributes format preserving encryption may
be an additional requirement for all of the listed elements.
\begin{table}[H]
\caption{\label{tab:Required-Properties-for}Required Properties for SQL}
\resizebox{\columnwidth}{!}{
\begin{raggedright}
\begin{tabular}{|l|c|c|c|c|}
\hline
 & deterministic & order & homomorphic & search\tabularnewline
\hline
\hline
RESTRICTION & X & X &  & \tabularnewline
\hline
PROJECTION  & X &  &  & \tabularnewline
\hline
JOIN & X &  &  & \tabularnewline
\hline
UNION & X &  &  & \tabularnewline
\hline
INTERSECTION & X &  &  & \tabularnewline
\hline
DIFFERENCE & X &  &  & \tabularnewline
\hline
SEMIJOIN & X &  &  & \tabularnewline
\hline
SEMIDIFFERENCE & X &  &  & \tabularnewline
\hline
EXTEND &  &  & X & \tabularnewline
\hline
AGGREGATE (AVG,SUM) &  &  & X & \tabularnewline
\hline
AGGREGATE (MIN,MAX) &  & X &  & \tabularnewline
\hline
SUMMARIZATION &  &  &  & \tabularnewline
\hline
ORDER BY &  & X &  & \tabularnewline
\hline
LIKE &  &  &  & X\tabularnewline
\hline
FUNCTIONS &  &  & X & \tabularnewline
\hline
\end{tabular}
\par\end{raggedright}}

\end{table}

%\newpage{}

\section{Order Preserving Encryption\label{sec:Order-Preserving-Encryption}}

As the requirements from the chapter before (\ref{subsec:Query-Requirements})
show, order preserving is an important property for many queries to
work. If sensitive data is encrypted order preserving, many queries
on the encrypted data are possible. As seen queries like ``order
by'' or ``>'' ``<'' in the where clause could work on the encrypted
data without changes. Other properties are dependent of the algorithm
used, in some cases order preserving includes also format preserving,
which means that all the advantages of format preserving encryption
can be applied here too. If there is a one to one mapping between
the plain text and the ciphertext, then even joins on the encrypted
fields are possible without change of the data model or the queries.
The drawback is that generally the more additional features an order
preserving algorithm supports, the less secure it is. Other features,
which require operations on the data like the aggregate functions
``Sum'' or ``Average'' are not supported directly. For theses
cases the whole data has to be decrypted before processing is possible.

Order preserving encryption can be seen also as a kind of property
preserving encryption. The property that should preserved is the order
relation. If the plain text value v1 > v2, then the encrypted value
v1' > v2' still holds. All order preserving ciphers presented here
are symmetric. The reason for this is that for the uses of order preserving
encryption, symmetric ciphers are more appropriate and not that asymmetric
order preserving ciphers are impossible. Order preserving ciphers
can have additional properties and constraints. Some of them work
only for specific datatypes like numbers or strings, while others
work for all. An order preserving cipher can be either deterministic
or probabilistic (one to many), which has also great impact on usability
and security. This section gives an overview of past and present order
preserving encryption schemes. Some of these schemes are described
more in detail than others, because they either had a significant
impact on the development of order preserving encryption, they are
the most practical or are the most secure currently available.

\subsection{Classical Schemes}

Some of the classical ciphers are already order preserving, or can
be easily made so. Although not practically from a security standpoint,
these ciphers can show some insight on order preserving encryption.
As they are easy to understand and implement, they can be used as
simple test bed for any solution using order preserving encryption.
Out of the box for example, Caesar cipher is not order preserving,
because the last letters of the alphabet are encrypted as the first.
For example a Caesar cipher with 3 as a key and only letters as the
alphabet. For the letters A - W it is order preserving, but for X,Y,Z
it is not because e(W) = Z and e(X) = A thus e(W) > e(X) is not correct.
To make it order preserving is easy, the range of the ciphertext is
extended. X,Y, Z is mapped to additional characters like ``.'',
``,'' and ``;''. Another way is to support only range queries.
Here the query is rewritten in such a way that the wrap-around case
is taken into account. This shows an interesting point: Knowledge
gained on such simple ciphers like the Caesar cipher can be used on
much more complicated and more secure ciphers. In this case, the solution
for range queries with the Caesar cipher is the same as for modular
order encryption.

\subsection{Modern Encryption Schemes}

Historically ciphers aside, it all started with the Hacigumus Scheme~\cite{Hacigumus:2002:ESO:564691.564717}, 
then OPES~\cite{Agrawal:2004:OPE:1007568.1007632},
which created the term Order Preserving Encryption (OPE). After that,
OPE~\cite{citeulike:9349465}, took a more formal approach to security,
defining ciphertext indistinguishability for order preserving algorithms.
Modular order preserving encryption~\cite{ope_revisited} was the
latest addition to order preserving ciphers, claiming the highest
security level, while still being practically usable. Besides these
ciphers many other encryption schemes were created in recent years.
These ciphers are only described here on the surface. Some of them
were only minor variations of existing ones, while other could not
deliver the security they promised. The research of order preserving
encryption is still very active, resulting in new schemes every year.
Also more knowledge of the security of theses ciphers is gained, but
still the security of order preserving encryption is not as well understood
as the security of standard block ciphers like AES. Thus most of the
researchers warn of the use of these ciphers in practical applications,
if the requirement for security is high. The following schemes are
all order preserving. This is a non-exhaustive list of existing schemes
and only an overview is given. For more details on these ciphers see
the original publications as stated in the bibliography.

\subsubsection{Bucket Based Approach}

The encrypted relation differs from the unencrypted relation. The
whole original tuple is stored in one attribute and for every searchable
attribute an index attribute is added. Each index attribute does not
contain the original value, but only the bucket value. Each bucket
is a subset of the attribute domain. Two strategies for selecting
the boundaries of the buckets are possible:
\begin{itemize}
\item equi-width. 
all buckets have the same range.
\item equi-depth. 
all buckets contain the same number of items.
\end{itemize}
The disadvantage of the equi-width strategy is that the distribution
of the attributes is revealed to the encrypted database. The downside
of the equi-depth strategy is, that data changes requires updates
of the bucket boundaries, so that all buckets still contain the same
number of items~\cite{Mykletun:2006:AQD:3127142.3127149}. To create
a bucket value the domain of each attribute is split in buckets, mapping
multiple values to one. To execute a query on the encrypted data the
values in the query are replaced by the corresponding bucket values.
After receiving the data, it is filtered on the client to remove the
spurious tuples.This makes queries on equality on the encrypted data
possible, but as the buckets are not sorted, range queries are, depending
on the domain either limited or not possible. Aggregation queries
are possible too. While using an homomorphic encryption scheme leaks
information as shown by~\cite{Mykletun:2006:AQD:3127142.3127149},
another approach is to predetermine aggregate functions like count
and sum for each bucket. These attributes are stored encrypted together
with the bucket id in the same way as normal index attributes~\cite{Hacigumus:2002:ESO:564691.564717}.

\paragraph*{Example}

\begin{table}[H]
\caption{Original Table (Bucket)}
\centering
\begin{tabular}{|c|c|c|c|}
\hline
CUSTOMERID & LASTNAME & AGE & INCOME\tabularnewline
\hline
\hline
1 & Miller & 22 & 30000\tabularnewline
\hline
2 & Smith & 54 & 90000\tabularnewline
\hline
3 & Hill & 21 & 25000\tabularnewline
\hline
4 & Moore & 77 & 35000\tabularnewline
\hline
\end{tabular}
\end{table}
The values are split to buckets:

\begin{lstlisting}
Lastname: H -> 8 M ->13 S -> 19
Age: 21 - 30 -> 3 51 - 60 -> 6 71 -80 -> 8
Income: 0 - 25000 -> B 25001 - 50000 -> C 50001 - 75000 -> D 75001 - 100000 -> A
\end{lstlisting}

\begin{table}[H]
\caption{Encrypted Table (Bucket)}
\centering
\begin{tabular}{|c|c|c|c|c|}
\hline
ID & TUPLE & I\_LAST & I\_AGE & I\_INCOME\tabularnewline
\hline
\hline
1 & \$\$\&\%\&/(\%/\&§'' & 13 & 3 & C\tabularnewline
\hline
2 & \%\$\%(\$\%§\$\&/§ & 19 & 6 & A\tabularnewline
\hline
3 & \&(\%\%/\$/(§''/\%/(/ & 8 & 3 & B\tabularnewline
\hline
4 & \&\%\&)\%()\%\%)\% & 13 & 8 & C\tabularnewline
\hline
\end{tabular}
\end{table}

\begin{table}[H]
\caption{Encrypted Aggregate Table (Bucket)}
%\resizebox{\columnwidth}{!}{
\centering
\begin{tabular}{|c|c|c|}
\hline
BUCKET\_ID & SUM & COUNT\tabularnewline
\hline
\hline
A & enc(90000) & 1\tabularnewline
\hline
B & enc(25000) & 1\tabularnewline
\hline
C & enc(65000) & 2\tabularnewline
\hline
D & enc(0) & 0\tabularnewline
\hline
\end{tabular}
%}
\end{table}

\paragraph*{Some example queries}

\paragraph*{Value query} 
client:

\begin{lstlisting}
SELECT * FROM CUSTOMERS WHERE LASTNAME = 'Miller';
\end{lstlisting}
server:

\begin{lstlisting}
SELECT TUPLE FROM CUSTOMERS WHERE I_LAST = 13;
\end{lstlisting}
client:

\begin{lstlisting}
decrypt each TUPLE, filter by LASTNAME 'Miller'
\end{lstlisting}

\paragraph*{Limited range query}
client:

\begin{lstlisting}
SELECT * FROM CUSTOMERS WHERE INCOME BETWEEN 20000 AND 40000;
\end{lstlisting}
server:

\begin{lstlisting}
SELECT TUPLE FROM CUSTOMERS WHERE I_INCOME = 'B' OR I_INCOME = 'C'
\end{lstlisting}
client:

\begin{lstlisting}
decrypt each TUPLE, filter by INCOME between 20000 and 40000
\end{lstlisting}

\paragraph*{Aggregate query}

client:

\begin{lstlisting}
SELECT SUM(INCOME) FROM CUSTOMERS
WHERE INCOME BETWEEN 20000 AND 50000;
\end{lstlisting}
server:

\begin{lstlisting}
SELECT SUM FROM AGG_INCOME WHERE BUCKET_ID = 'C';
SELECT TUPLE FROM CUSTOMERS WHERE I_INCOME = 'D';
\end{lstlisting}
client:

\begin{lstlisting}
decrypt all tuples, filter by INCOME >= 20000, calculate sum
decrypt sum from AGG_INCOME and add calculated sum
\end{lstlisting}

\paragraph*{Hash Based Approach}

This approach is very similar to the bucket approach but instead of
splitting the domain of an attribute into buckets a one-way hash function
is applied. If the hash function is not collision free then the spurious
tuples have to be filtered in this approach, too. Queries on equality
are supported, but range queries are not possible, because the hashes
do not preserve the order of the original values.

\paragraph*{B+ Tree Approach}

This requires a more different schema than the previous approaches.
Here the data is stored as a b+ tree. A vertex has a id and the content,
which contains references to the lesser and to greater vertex and
leaf nodes contain the links to the other nodes. This content is encrypted
and to query the data, the tree has to be traversed by the decrypted
vertexes until an leaf node is found. Although more steps for the
retrieval are necessary, this makes range queries possible.

\subsubsection{OPES (Order Preserving Encryption Scheme)}

The intuition behind this algorithm~\cite{Agrawal:2004:OPE:1007568.1007632}
is the following: Values from a user-specified distribution are generated,
and sorted in a table. The index of an value in this table is the
encrypted value. This table is the key. Decryption is a simple lookup
in this table, which has the role of the encryption/decryption key.
So the only thing revealed by the encrypted value is the order, which
is exactly what is wanted. In reality, this is not practical, because
the key is large, and every update can require a complete encryption.
The goal of OPES is to construct an encryption function which has
the same properties. OPES is the first order preserving encryption
scheme which is constructed with the relational data model in mind. OPES works in three stages:
\begin{enumerate}
\item Model. 
Input and target distributions are modeled as piece-wise linear
splines. Here the data values are partitioned into buckets, and each
bucket is as linear spline.
\item Flatten. 
The plain text values are transformed, so that the values
are uniformly distributed. Values from a bucket are mapped to buckets
with length proportional to the number of values.
\item Transform. 
The flattened values are mapped into the target distribution
and then encrypted.
\end{enumerate}

\subsubsection{OPE (Order Preserving Encryption)\label{subsec:OPE-(order-preserving}}

\paragraph*{Hypergeometric probability function}

There is a relation between a random order preserving function and
hypergeometric probability distribution. Any order preserving function
f from \{1,....M\} to \{1,....N\} can be represented by a combination
of M out of N ordered items. This can be represented by a bin with
N balls. M balls are black and N-M balls are white. The at each step
a random ball without replacement is drawn. The random variable X
is the number of black balls in the sample, after collecting the y-th
ball. This variable has a hypergeometric distribution. The probability
of X=x is given by
\[
\frac{\left(\frac{x}{y}\right)\left(\left(\frac{N-y}{M-x}\right)\right)}{\left(\frac{N}{M}\right)}
\]
If the y-th ball is black, then the least unmapped point of the domain
is mapped to y. A simple example with a very small domain and range
could be: domain D: {[}1,2,3{]} which is mapped to R: {[}1,2,3,4,5,6,7,8,9{]}.
The experiment would be to withdraw random balls from the set. In
this example the set would contain 3 black balls (elements from D),
and 6 white (elements from R) {[}{*},{*},{*},o,o,o,o,o,o{]}. A possible
random sequence of chosen balls could be \{o,o,{*},{*},o,{*}\}. The
following steps are performed:
\begin{enumerate}
\item {[}{*},{*},{*},o,o,o,o,o{]} y = 1, o is chosen.
\item {[}{*},{*},{*},o,o,o,o{]} y = 2, o is chosen.
\item {[}{*},{*},o,o,o,o{]}, y = 3, {*} is chosen the least unmapped number
of D is mapped {[}1 -> 3{]}.
\item {[}{*},o,o,o,o{]}, y = 4, {*} is chosen, the least unmapped number
of D is mapped {[}2 -> 4{]}.
\item {[}{*},o,o,o{]}, y = 5, o is chosen.
\item {[}o,o,o{]}, y = 6, {*} is chosen, the least unmapped number of D
is mapped {[}3 -> 6{]}.
\end{enumerate}
As this is not efficient for real domains, a more efficient function
is given by~\cite{citeulike:9349465}, which recursively samples a
random order-preserving function. This makes the solution more practical,
although the rationale stays the same~\cite[pp 3]{citeulike:9349465}.

\paragraph*{Multiple messages and state}

As the independent encryption of multiple plaintexts would not be
ordered, and keeping state of all encrypting messages is cumbersome,
the state can be assumed as an static, but random tape. To encrypt
plain text x, the encryption algorithm performs a binary search of
x by recursively calling the encryption with e(K,M/4) if m < M/2 or
with e(K,3M/4) if greater. ``Each ciphertext is made out of the hyper-geometric
sampling algorithm and coins from an associated portion of the random
tape, indexed by the plain text''. A pseudo-random function can be
created, to generate the tape dynamically. This function should be
block cipher-based~\cite[pp 3]{citeulike:9349465}.

\paragraph*{Encryption and decryption}

D = Domain \{1..M\}, R = Range \{1..N\} plain text m \ensuremath{\in}{[}D{]},
ciphertext y \ensuremath{\in}{[}R{]} A Range gap is mapped to a domain
gap. The algorithm is called with sets of the domain and the range.
The start value for the range gap (y) = N/2. After generating pseudo-random
coins and giving them to the hypergeometric sampling function x is
calculated. The number of values of the order preserving function
which are less than y is the domain gap x. The mth point of the ciphertext
is m. If m is less than x the encryption function is recursively called
with the subset D\{d+1,x\} and R\{r+1,y\} or if greater with D\{x..d+M\}
and R\{y..r+N\}. If the number of the domain is 1 the algorithm ends
returning the ciphertext by choosing a point from the set as result.The
decryption is very similar. At the end of the recursion if the ciphertext
is in the range, the message m is returned as result. It is crucial
for the algorithm to choose the right size of N. The goal is to have
a large number of random order preserving functions. It is suggested
by the authors, that N = 2M, which results in more than 2\textasciicircum 80
functions.

\begin{figure}[ht]
\centering
\includegraphics[width=\columnwidth]{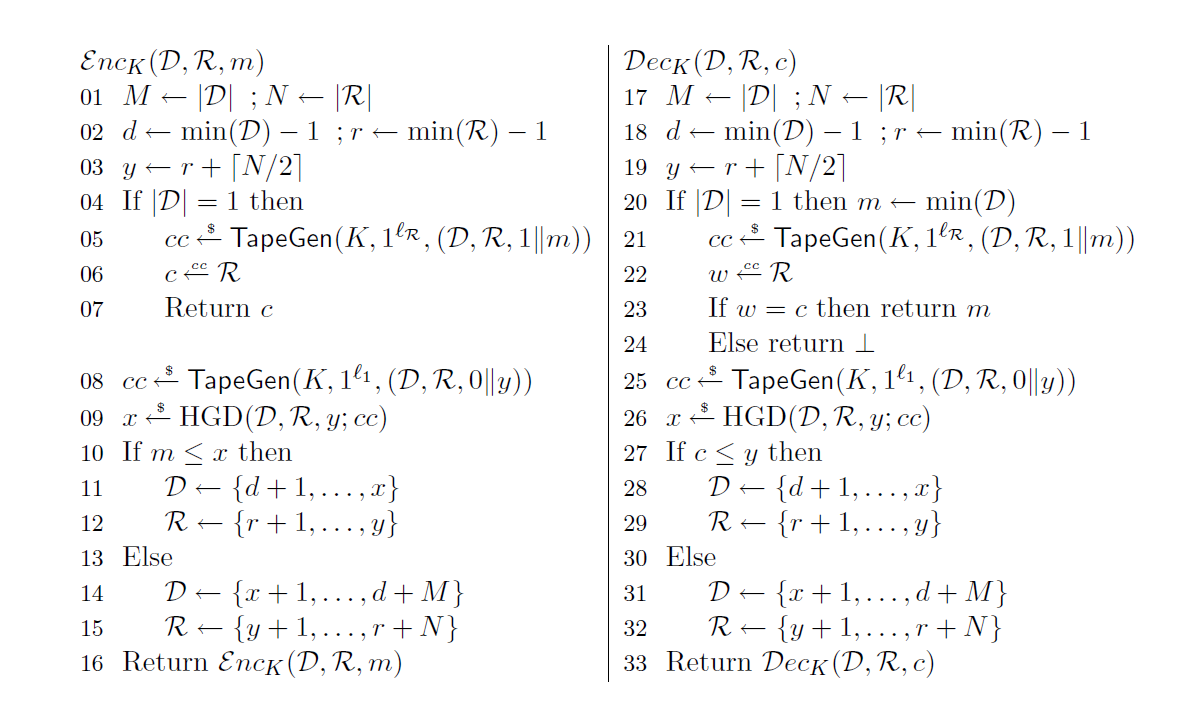}

\caption{OPE Algorithm~\cite{citeulike:9349465}}
\end{figure}

As already mentioned, order preserving encryption (OPE) is a encryption
where the order of the cleartext is retained in the cipher text. It
is crucial that only the order and no other information of the clear
text is revealed in the cipher text.It is a kind of homomorphic encryption,
where the homomorphic operation is order comparison~\cite{Agrawal:2004:OPE:1007568.1007632,citeulike:9349465,Mengke_2010}.

\subsubsection{MOPE (Modular Order Preserving Encryption)}

Modular OPE is a modification of order preserving encryption~\cite{citeulike:9349465}.
It adds a secret modular offset to the plain text before encryption.
The scheme is not strictly order preserving anymore, but it permits
range queries. For the execution of a modular range query two cases
have to to be considered: The standard case occurs, if the range values
are ordered c1 < c2. For this case, there is no difference between
the execution on the plain text and the ciphertext. In the second
case (c2 > c1), also called the wrap around case, the query on the
ciphertext differs: Here the query is executed with range {[}c1, M{]}
and range {[}1,c2{]}. Consider the following simple example: D = {[}1,2,3,4,5,6{]}
M = 5, R = {[}1,2,3,4,5,6,7{]}, j = 2, the trivial OPE f(x) -> x +
1. The data is {[}1,2,2,3,4,5,5,5{]} the encrypted data is {[}4,5,5,6,1,2,2,2{]}.
\begin{itemize}
\item Standard query. 
select t.{*} from table t where t.a1 between 1 and
3. The result is {[}1,2,2,3{]}. The offset is 2 so the lower bound
1 is encrypted as (1 + 2 mod 6) = enc(3) = 4. The upper bound 3 is
encrypted as (3 +2 mod 6) = 5 = 6. As 4 < 6 this is the standard case
and the encrypted query is select t.{*} from table t where t.a1 between
4 and 6. The encrypted result is {[}4,5,5,6{]} which is decrypted
to {[}1,2,2,3{]}.
\item Wrap around query. 
select t.{*} from table t where t.a1 between 3
and 5. The offset is 2 again. The lower bound 3 is encrypted as (3
+ 2 mod 6) = enc(5) = 6 and the upper bound is (5 + 2) mod 6 = enc(1)
= 2. As 6 is not < 2 the query is rewritten to: select t.{*} from
table t where t.a1 between 6 and 7 union select t.{*} from table t
where t.a1 between 1 and 2. The encrypted result is {[}6{]} union
{[}1,2,2,2{]} which is decrypted as {[}3{]} union {[}4,5,5,5{]}~\cite{ope_revisited,Mavroforakis:2015:MOE:2723372.2749455}.
\end{itemize}

\subsubsection{MV-POPES (Multivalued-Partial Order Preserving Encryption Scheme) }

The domain of the plain text is divided in multiple partitions which
are randomized in the encryption domain.To enhance the security of
OPE this scheme encrypts an integer to multiple different values.
In each partition the encrypted values are ordered.Range queries have
an higher overhead than in OPE. The overhead depends on the number
of the partitions, but can be reduced with multilevel partitioning
or binary recursive partitioning~\cite{mv-popes}.

\subsubsection{mOPE (Mutable Order Preserving Encoding)}

This scheme is more a protocol, than a cipher. According to~\cite{popa_modular}
it is defined as: ``A mutable order-preserving encryption scheme
for plain text domain D is a tuple of polynomial-time algorithms mOPE=
(KeyGen, InitState, Enc, Dec, Order) run by a client and a stateful
server, where KeyGen is probabilistic and the rest are deterministic,
and Enc is interactive.`` It requires state on a server, and this
state has to be mutable. Although more complicated than other order
preserving solutions, it achieves the highest level of security for
order preserving encryption IND-OPCA. This means that only the order,
but nothing else is revealed about the encrypted values. The values
are ciphered on the client with a symmetric cipher and stored in a
search tree on the server. The tree traversal for a value v is performed
by these steps:
\begin{figure}[ht]
\centering
\includegraphics[width=\columnwidth]{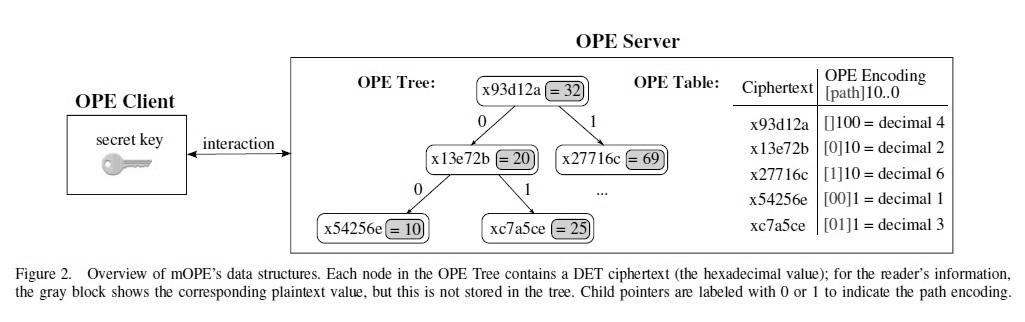}

\caption{mOPE data structures~\cite{popa_modular}}
\end{figure}

\paragraph*{Request root from server}
\begin{enumerate}
\item Decrypt value v on client as v', if v < v' request left, if v' > v
right from the server, if v = v' request found.
\item Repeat step 2 until v is found or empty returned.
\item The result is the path from the tree and whether the value was found.
\end{enumerate}

\paragraph*{To insert a new value, the following steps are performed:}
\begin{enumerate}
\item Encrypt value v on the client.
\item OPE tree is traversed (as in the 1st algorithm), and inserted, possible
balanced.
\item To query, the server finds the order in the tree of the encrypted
value or the bounds and executes the query on the encrypted data using
an user defined function (order).
\end{enumerate}

\subsubsection{DOPE (Dynamic Order Preserving Encryption)}

This scheme is a enhancement of mOPE. The main difference is, that
the OPE tree is stored in the database by using an AVL tree~\cite{dope}.

\subsubsection{FH-OPE (Frequency Hiding Order Preserving Encryption)}

This scheme keeps local state of the plain text and the encrypted
order. Only the ciphertext is sent to the server.The encryption is
key-less, the security relies on the state of the algorithm. An update
can potentially affect the complete ciphertext.It assumes uniformity
of distribution of the plain text.For decryption only a lookup in
the state tree is required.This scheme achieves better performance
than the IND-OPCA Scheme, but the assumption of uniformity of the
distribution is a significant constraint~\cite{Kerschbaum:2014:OAI:2660267.2660277}.
\begin{itemize}
\item IND-OPCA Scheme. 
This scheme requires statefulness.The state is encrypted
in a tree data structure at the server and therefore the encryption
process requires multiple requests between the client and the server.
On the server the search tree is used to query the encrypted data.
Additionally the complete ciphertext has to be updated after some
time for not leaking the distribution of the plain text.If the plain
text has uniform distribution, than no updates are required, but in
reality, this is not often the case~\cite{10.1007/978-3-319-78556-1_3,frequ_ope}.
\end{itemize}

\subsubsection{COPE (Chaotic Order Preserving Encryption) }

This cipher works only on a trusted database, as the data is decrypted
and encrypted on the server. The transformation of the plain text
values is performed in two steps: The goal of the first step is to
hide the order of the plain text. The second step preserves the order
to allow efficient queries.
\begin{itemize}
\item Random Shuffling. 
The domain of the data is partitioned in buckets.
Each of these buckets has either descending or ascending order and
the buckets are randomly mixed. To partition the domain, a number
of random values is generated, where the order of the bucket is determined
by the sequence. The order within the bucket can be either ascending
or descending and is determined by the number of the bucket.
\item Chaotic Beta-Expansion. 
The randomly shuffled database is transformed
into an ordered ciphered database by using Beta-expansion.The encrypted
values preserves the lexicographic order. This makes fast queries
on the encrypted database possible~\cite{cope}.
\end{itemize}

\subsubsection{SOPE (Semi-Order Preserving Encryption)}

Although similar to order preserving encryption the resulting ciphertext
is not strictly ordered, but only ordered to some degree. It is possible
to encrypt two different values to the same ciphertext, which is not
possible in OPE. If the encryption function is f(x) -> y and x1 <
x2, then only the constraint y1 <= y2 and not y1 < y2 is satisfied.
The ciphertext space of SOPE can be divided in two sets: One that
satisfies the order preserving condition, the other does not satisfy
the condition. The ratio between these two sets measures the difference
between OPE and SOPE. If all values are in the ordered set, it is
in fact an OPE.The degree d is the sum of the probability of plain
text encoded in the same value. A degree of 0 means that every plain
text is encoded in a different ciphertext, while 1 means that at least
2 plaintexts are mapped to the same ciphertext.The higher the semi-order
preserving degree is, the better is the security of the cipher. The
drawback is that the error rate is higher, too. As different plaintexts
are mapped to same values, the encryption has to keep state to make
it possible to decrypt the value to the original value again~\cite{YANG2017266}.

\subsubsection{p-OPE (Probability-p Order Preserving Encryption)}

This is similar to the semi-order preserving encryption scheme. The
resulting ciphertext is ordered, but only to a certain degree. Not
all ciphertext values are in the correct order. If a query is executed
on the encrypted data, the result can contain false positives (values
which are not in the result of the query on the plain text) and it
can even contain not all queried values (false negatives). The result
of an query on encrypted data is not the same as the query on the
cleartext data.The deviation of the correct result depends on the
probability and it can be acceptable to loose some precision to gain
more security~\cite{Yang20182PO}.

\subsubsection{New order preserving encryption model}

This OPE model sets the focus on outsourced databases in cloud environments.
The goal is to be more secure than OPE, but also to avoid the performance
penalty of secure OPE like IND-OCPA.The encryption scheme mitigates
statistical attacks (the data distribution/ data frequency), which
are the weak spot for OPE. To achieve this, the message space is extended.
This can be done by two different ways. One way is to keep the type
of the attribute but increase the precision. For example a real (8,4)
can be enhance to a real(12,8). The other, more invasive way is to
represent the number as string. The encryption is performed in different
steps: 1) The message space is split. This destroys the data distribution.
2) The ciphertext space is split. 3) Each value is mapped to the ciphertext
space. Both the encryption and decryption is run on the client~\cite{LIU2016198}.

\subsubsection{Order-Preserving Encryption Using Approximate Integer Common Divisors}

As other OPE schemes the ciphertext has to be significant larger than
the plain text space.The security of this scheme is given by the general
approximate common divisor problem. For additional security, t he
scheme can be used in conjunction with other OPE schemes like the
original OPE~\cite{DBLP:journals/corr/DyerDX17a}.

\subsubsection{One-to-Many OPE}

Instead of mapping a plain text value to an encrypted value, each
value is mapped to a bucket, and from this bucket a encrypted value
is chosen randomly. This requires a significant larger ciphertext
space, but the frequency of the plain text is hidden, because it is
mapped to different values.Although the distribution is effectively
hidden, an differential attack is still possible. Common values are
mapped to the same bucket, which means that from this bucket many
values are used, with little difference between them~\cite{7110579}.

\subsubsection{NOPE (Noise Based Order Preserving Encryption)}

This encryption scheme generates a noised based encryption function
for enhanced security~\cite{ope_sec_analysis}.

\subsubsection{sOPE (Stateful Order Preserving Encryption)}

The encryption algorithm is separated between an order preserving
and a symmetric key encryption part. A ciphertext c contains (c1,c2)
where c1 is the ordered and c2 is the symmetric encrypted part. For
c1, a key point is that the the ratio of the partitions in the plain
text and ciphertext space are the same.All values are stored in a
table, and for encryption each value is looked up in this table.This
table can be stored either on the client or on the server. For decryption
only c2 is needed~\cite{10.1007/978-3-319-78556-1_3}.

\subsubsection{TOPE (Top Order Preserving Encryption)}

This scheme is for retrieving the top relevant tuples, also called
top-k queries. A simple example is the query ``select max(a) from
table'', which selects the max value of a column of a table . Another
would be ``select {*} from table order by attr desc limit 10'',
which selects the 10 tuples where a has the highest value. These queries
can be executed over the encrypted data without revealing any other
information about the data (even the ordering of the non top-k data).
The security is defined and proved with ``indistinguishability under
top-ordered chosen plain text'' attacks. The schemes utilizes a partially
ordered tree structure (heap) for min and max values. The values themselves
are encrypted with a standard symmetric encryption cipher. While the
encryption of the value itself never changes (permanent ciphertext),
the top/min information has to be updated (transient ciphertext),
if additional data is created, updated or deleted. The state of the
heap is maintained on the server, while the encryption/decryption
is done on the client. The state could be kept on the client too,
but this would require a lot of local storage on the client, and if
used on multiple clients multiple copies of the heap would exists.
On the other hand, the key on the server could compromise security.
A query is executed on the concatenated values of the transient and
permanent ciphertext, and the result, the permanent value and or other
attributes are retrieved and decrypted on the client~\cite{8385101}.

\subsection{Security of Order Preserving Encryption}

Pseudo Random Permutation (PRP) and Pseudo Random Function (PRF) are
security notions. A common approach it the ``Rank-then-Encipher Approach'':
At first the rank of each plain text message is calculated, so that
each plain text message has a rank. This rank is encrypted. The plain
text message is replaced with the plain text message of the encrypted
rank. To decrypt the rank of the decrypted message is calculated and
encrypted. The message is replaced with the plain text of the decrypted
rank~\cite{cryptoeprint:2009:251,Rogaway10asynopsis}. It
is no secret, that security is the most important quality of a cipher.
A cipher without security is pretty useless. Of course an unusable
secure cipher is useless, too. So there is always an area of tension
between security and usability. Ciphers without constraints or required
properties like order or format are always potentially more secure
than property preserving algorithms, because the attack surface is
smaller. Another point is that standard (not property preserving)
ciphers are much better researched and analyzed, because they are
much more common and widespread than order preserving ciphers.

\paragraph*{ROPE (Random Order Preserving Encryption)}

This is just a theoretical order preserving encryption scheme for
comparison with a real one. It is used to describe an ideal order
preserving encryption scheme. It is used in security analysis, to
analyze which information is leaked from an order preserving encryption
scheme~\cite{ope_revisited}.

\subsubsection*{Ciphertext Indistinguishability}

Ciphertext Indistinguishability is a property of an encryption scheme.
It means that an adversary cannot distinguish the ciphertext of two
encrypted messages with the same length. If an adversary can distinguish
the ciphertext with a probability significant greater than 0.5, the
the encryption scheme is not secure in the terms of indistinguishability.(wiki)
The type of attack is specified for the distinguishability like indistinguishability
under chosen plain text (IND-CPA), chosen ciphertext (IND-CCA) or
another kind of attack.

\paragraph*{Definition CPA Security}

Let HE = \{Gen,Enc,Dec,Eval\} be an encryption scheme. Given any adversary
A, we consider an experiment between the adversary A and the following
challenger C:
\begin{itemize}
\item C runs (pk, sk) \textleftarrow{} Gen(1\textgreek{k}) and sends pk
%\item C runs (pk, sk) \textleftarrow{} Gen(1\textkappa) and sends pk
to the adversary A.
\item A selects two messages in the message space M, i.e., (m0 ,m1)M\texttimes M,
and sends them to C.
\item C flips a uniformly random bit b \ensuremath{\in} \{0, 1\} and sends
c{*} \textleftarrow{} HE.Enc(pk, mb) to A. Here c{*} is called the
challenge ciphertext.
\item A outputs b\textquoteright{} \ensuremath{\in}\{0, 1\}. We say A wins
the game if b\textquoteright{} = b, i.e., A correctly finds out the
bit b~\cite[pp 334]{Vacca:2016:CCS:3098688}.
\end{itemize}
Indistinguishability under chosen plain text attack (IND-CPA), also
known as polynomial security is a strong security property for (asymmetric)
encryption algorithms. Of course this can never be achieved by an
order preserving encryption scheme, because as the order in the ciphertext
is preserved, it is always possible to distinguish the ciphertext
by simply comparing the order of the ciphertext. Indistinguishability
under chosen ciphertext attack/adaptive chosen ciphertext attacks
(IND-CCA1 / IND-CCA2) are even stronger security properties. Here
additional to IND-CPA the adversary has access to the decryption oracle,
meaning the adversary can encrypt and decrypt messages. IND-CCA1 restricts
the access to decryption oracle to point where the two messages, which
should be distinguished are received, while IND-CCA2 can access the
encryption oracle even after receiving the messages. Of course it
is not allowed to use the oracle on the messages received. Other weaker
models exist too: IND-DCPA (Indistinguishability under distinct chosen
plain text attack) for example is a weakened IND-CPA, requiring the
adversary to only choose distinct messages. For order preserving encryption
the following ciphertext indistinguishability models exist
\begin{itemize}
\item Indistinguishability under ordered chosen-plaintext attack (IND-OCPA).
This is the ideal security model. No information other than the order
is revealed.
\item Indistinguishability Under Frequency-Analyzing Ordered Chosen Plaintext
Attacks (IND-FA-OCPA). 
It is an generalization of IND-OCPA. According
to~\cite{frequ_ope} it is not achievable, but an achievable definition
IND-FA-OCPA{*} is suggested. Because of security concerns this kind
of OPE is not recommended.
\item Pseudo-random order-preserving function, security under chosen-ciphertext
attack (POPF-CCA). 
Here the oracle access to the encryption algorithm
of the order preserving encryption function and a random ordered encryption
function are indistinguishable under chosen ciphertext attacks. This
security model is weaker than IND-OCPA. It is proven, that at least
half of the bits of a plain text are leaked. According to Popa~\cite{popa_modular} this
goal can only be achieved by statefulness on the client and ciphertext
mutability. Ciphertext mutability means that when a new ciphertext
value is added, existing ciphertext is updated~\cite{10.1007/978-3-319-78556-1_3}.
\item \textgreek{j}-lsb-KPA. 
In this model the secrecy of the fraction (\textgreek{j})
of the least significant bits is guaranteed under known plain text
attacks.
\item \textgreek{d}-IND-OCPA. 
A stronger security goal than \textgreek{j}-lsb-KPA
and for \textgreek{d} = 1 even IND-OPCA~\cite{10.1007/978-3-319-78556-1_3}.
\item random order-preserving function (ROPF). 
A random order preserving
function is seen as an ideal object. If an order preserving function
cannot not distinguished from this ideal object indistinguishability
to ROPF is achieved~\cite{ope_revisited}.
\end{itemize}

\paragraph*{Security Notions for Order Preserving Encryption}
\begin{itemize}
\item One-Wayness. 
The is a very fundamental security requirement. It states
that it is not possible to recover the message from a ciphertext without
the key. Of course this notion is not as strong as it seems, because
even if the cleartext itself is not known, other information may be
revealed~\cite{cryptoeprint:2015:1125}.
\item Window One-Wayness. 
This is a stronger security notion. The adversary
is successful if the message is contained in an interval of cleartext.
As an example the value 1000 is encrypted. If the given interval is
100 and the adversary finds the value is 960, it counts as successful
decrypted~\cite{ope_revisited}.
\item Window distance One-Wayness. 
This metric tries to quantify if the information
of the distance between numbers of the plain text are revealed. For
example if values 5 and 6 are ciphered, the order preserving encrypted
values should reveal that e(6) > e(5) but not that the distance between
e(5) and e(6) is 1.
\end{itemize}

\paragraph*{Security Models:}

Leakage profiles show if and how much additional information is revealed
by an encryption scheme. The following profiles exist~\cite{Durak:2016:ERO:2976749.2978379,cryptoeprint:2015:1125}:
\begin{itemize}
\item Ideal. 
Only the order of the ciphertexts is revealed. It hides any
statistical information about the gaps between the messages.
\item ROPF. 
Random order-preserving function: As later shown reveals at
least half of the plain text bits.
\item MSDB. 
Most-significant-differing bit profile: The most significant
bit are allowed to leaked.
\item TrM, MtR are other leakage profiles mentioned in literature~\cite[pp 4]{Durak:2016:ERO:2976749.2978379}.
\end{itemize}

\subsubsection*{Attack models:}
\begin{itemize}
\item Known ciphertext model. 
The attacker can only access the encrypted
files without any additional background information.
\item Known background model. 
The attacker has not only access to the encrypted
data but also additional information like statistical information
about the data.If known which kind of data is stored, this can be
used for statistical attacks.
\end{itemize}
Attacks on order revealing encryption~\cite{Durak:2016:ERO:2976749.2978379}:
\begin{itemize}
\item Inter-column correlation-based attacks. 
As columns are often correlated,
an attacker can attempt to reveal more information from multiple order
revealing encrypted columns.
\item Inter + intra-column correlation-based attack.
\end{itemize}
The following example shows the result of an attack on order revealing
encryption. Although not one exact value is revealed, a lot insight
to the data can be gained.

\begin{figure}[ht]
\centering
\includegraphics[width=\columnwidth]{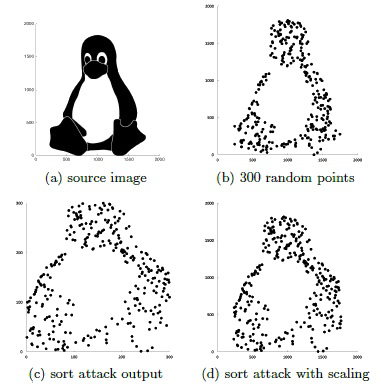}

\caption{\label{fig:2D-Attack}2D Attack~\cite[pp 5]{Durak:2016:ERO:2976749.2978379}}
\end{figure}

\subsubsection*{Attack Scenario Order Preserving Encryption: Information Leakage}

If an attacker can insert new entries in the database via the application,
it is possible that the order preserving encryption leaks information.
In this simple example a row exists with salary data of an employee.
Then the attacker creates a new record with the application, now he
only has to query the (encrypted) database with a simple query like
``select {*} from employee order by salary desc'' to know, if the
target earns more than the created dummy employee. This can be continued
until the exact salary is disclosed.

\begin{table}[H]
\caption{Information Leak OPE}
\begin{tabulary}{\columnwidth}{|c|C|C|}
%\begin{tabular}{|c|c|c|}
\hline
ID & Description & Salary (order preserving encrypted)\tabularnewline
\hline
\hline
1 & The target of the attack & 9897899\tabularnewline
\hline
2 & Dummy Row with salary €5000 (created by attacker) & 7866688\tabularnewline
\hline
3 & Other Rows & \tabularnewline
\hline
... &  & \tabularnewline
\hline
\end{tabulary}
%\end{tabular}

\end{table}

\subsection{Summary of Security of Order Preserving Encryption}

In~\cite{popa_modular} order preserving encryption schemes were analyzed
regarding leakage besides order and any security guarantees given.
To complete this chapter an overview of security and leakage beside
order of some of the ciphers shown before, is given:

\begin{table}[H]
\caption{Excerpt from comparison of ciphers in~\cite{popa_modular}}
\resizebox{\columnwidth}{!}{
\begin{tabular}{|c|c|c|}
\hline
Order-preserving scheme & Guarantees & Leakage besides order\tabularnewline
\hline
\hline
Agrawal et al.\textquoteright 04~\cite{Agrawal:2004:OPE:1007568.1007632} & None & Yes\tabularnewline
\hline
Boldyreva et al.~\cite{citeulike:9349465} & ROPF~\cite{citeulike:9349465} & §II-A Half of plain text bits\tabularnewline
\hline
Lee et al.\textquoteright 09~\cite{cope} & None  & Yes\tabularnewline
\hline
Yum et al.\textquoteright 12 {[}40{]}  &  ROPF~\cite{citeulike:9349465} & §II-A Half of plain text bits \tabularnewline
\hline
Popa, et.al~\cite{popa_modular} & IND-OCPA & None\tabularnewline
\hline
\end{tabular}}
\end{table}

%\newpage{}

\section{Property Preserving Encryption\label{sec:Property-Preserving-Encryption}}

Aside from order preserving encryption schemes there exist encryption
schemes, which preserve other properties of a plaintext. Order preserving
encryption, order revealing encryption, format preserving encryption,
searchable encryption, prefix preserving encryption and even homomorphic
encryption schemes can all be classified as property preserving encryption
schemes. All of the following encryption schemes are property preserving
encryption schemes. A property preserving cipher is optimal, if a
certain property of the plain text remains in the ciphertext without
revealing any additional information. At the end homomorphic encryption
is presented. Homomorphic encryption makes it possible, to perform
some operations like ``plus'' or ``multiplication'' on the encrypted
data without decrypting it first. Homomorphic encryption is neither
order preserving nor format preserving, but as arbitrary functions
can applied on the encrypted data, many useful properties can be implemented
as functions. The cost of these operations is multiple magnitudes
bigger than the operations on the plain text data, so although a very
promising approach for future research, it is not ready for use for
now.

\subsection{FPE (Format Preserving Encryption)}

Ciphertext normally does not meet the format constraints of an existing
data model. Fuzzy queries, range queries and other SQL operations can
not simply executed over encrypted data. The usage of symmetric ciphers
like AES results in binary strings regardless of the type or format
in the database or application. Therefore the data model and the application
has to be changed. To avoid this, FPE preserves data type and length
in ciphertext, so the existing schema of the database is still usable.
Other characteristics of the data like order preserving, logical laws
or domain specific checks are although possibly lost. All queries
are the possible on a technical level, but range queries for example
would not give the correct result, because operators like ``>'',
``<'' are not be valid for the encrypted data anymore. For such
queries, the whole data has to be retrieved, decrypted and then processed,
which can lead to a big overhead. If there are no such restrictions
on the sensitive data fields, it can be a good solution for legacy
systems, because the existing data model can be used without modification.
The quality of the encryption is not only dependent on the algorithm
but also by the length of the domain. Depending on the format, it
is usually not as secure as the use of standard block ciphers like
AES.

\paragraph*{Schemes }

According to~\cite{Rogaway10asynopsis}, format preserving ciphers
can be be split based on the size of the space.
\begin{table}[H]
\caption{\label{tab:taxonomy-format-preserving}Taxonomy format preserving
encryption~\cite{Rogaway10asynopsis}}
\centering
\begin{tabular}{|c|c|c|}
\hline
setting & size & msg space\tabularnewline
\hline
\hline
tiny-space FPE & N \ensuremath{\le} 2\textasciicircum 10 & X = {[}N{]}\tabularnewline
\hline
small-space FPE & N \ensuremath{\le} 2\textasciicircum 128 & X = \ensuremath{\sum}n\tabularnewline
\hline
large-space FPE & N \ensuremath{\ge} 2\textasciicircum 128 & X = \{0,1\}n\tabularnewline
\hline
\end{tabular}
\end{table}
As seen in table \ref{tab:taxonomy-format-preserving} format preserving
ciphers can be categorized by size and space. If N is sufficiently
small, then it is acceptable to spend O(N) time for key setup or the
first encryption. This is the easy case. If the message space is small
like X = \ensuremath{\sum}n for some arbitrary alphabet \ensuremath{\sum},
the a solution can often be based on Feistle networks. FFX is an example
for such small-space FPE. The last category is large-space FPE. It
is also called wide-block encryption. Standards are EME2 and XCB.
For tiny spaces some simple and provable secure format preserving
encryption schemes exist:

\paragraph{Knuth shuffle (Fisher Yates shuffle)}

Shuffle a set of numbers {[}x1..Xt{]}~\cite[pp 145]{Knuth:1997:ACP:270146}.
\begin{itemize}
\item P1. Initialize, Set j <- t,
\item P2. Generate a random number U uniformly distributed between zero
and one.
\item P3. Exchange Set k <- {[}jU{]} +1, exchange Xk <-> Xj.
\item P4. Decrease j. Decrease j by 1 if j > 1, return to step P2.
\end{itemize}

\paragraph{Permutation numbering}

Map key K to a number k e {[}N!{]} and encrypt with the kth permutation
on {[}N{]}~\cite{Rogaway10asynopsis}.

\paragraph{Prefix cipher }

Use ordering of Ek(0)..Ek(N-1) to determine permutation. AES can be
used for the encryption, to determine the order. As an output, a sorted
table is created. This table can be used for table-driven encryption~\cite{Rogaway10asynopsis}.

\paragraph{FFX Algorithms}

NIST Special Publication 800-38G, ``Recommendation for Block Cipher
Modes of Operation''~\cite{Dworkin2013} specifies some methods for
format preserving encryption. This publication contains two recommendations
for algorithms, which can be used for legacy applications, where the
format can not be changed or to generally encrypt sensitive information
while keeping the format. Both of them are based on a block cipher.
Currently AES with key lengths of 128, 192 and 256 bits are supported.
The main aspect of format preserving algorithms is that data is not
necessarily binary and the encrypted text may not have the same length
as the plain text. The alphabet (allowed symbols of the data) can
be restricted, for example only numbers or certain letters could be
allowed. FF1 and FF3 are the two recommended modes. They work like
this: Each symbol is represented by a numeral, base is the number
of the alphabet, denoted as radix. Two related function exist for
encryption and decryption. For encryption, the input is the plain
text given as numerical string (X) and a byte string as tweak (T),
the resulting output is a numerical string Y of the same length as
X. The decryption function has as input the encrypted numerical string
and the tweak and as output the plain text as numerical string. For
the same tweak the decryption function is the inverse of the encryption
function. A tweak does not have to be a secret, it can be data associated
to the plain text. Although not mandatory variable tweaks enhance
the security, because the chance of identical encryptions of the identical
values is reduced. Both algorithms, FF1 and FF3 are based on the Feistel
structure (\ref{fig:Feistel-Structure}). The Feistel structure consists
of several iterations, called rounds, of a reversible transformation.

\paragraph*{Outline of FF3}
\begin{itemize}
\item Split data in two parts.
\item Apply a function on one part of the data to modify the other part.
\item Swap roles for the next round.
\begin{figure}[ht]
\centering
\includegraphics[width=\columnwidth]{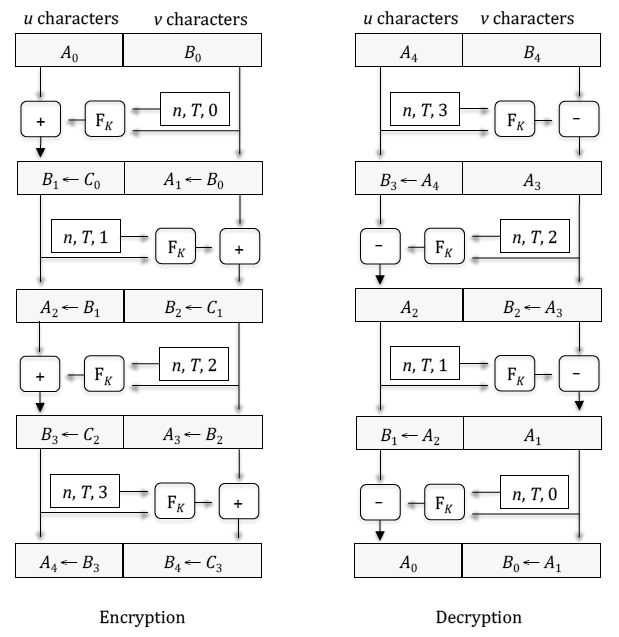}

\caption{\label{fig:Feistel-Structure}Feistel Structure~\cite[pp 10]{Dworkin2013} }
\end{figure}
\end{itemize}
In \ref{fig:Feistel-Structure} four rounds of the transformation
are shown.

\subsubsection{Length Limits of Modes}

\subsubsection{FF1 \label{subsec:FF1}}

The allowed parameters for FF1 are: radix: {[}2..2\textasciicircum 16{]}
and radix\textasciicircum minlen > 100, but much higher values are
recommended resulting in the range of 2 <minlen < maxlen < 2\textasciicircum 32.

\subsubsection{FF3}

The allowed parameters for FF3 are radix: {[}2..2\textasciicircum 16{]}
and radix\textasciicircum minlen > 100 and the possible values are:
2 < minlen < maxlen < 2{[}log.radix(2\textasciicircum 96){]}.

\subsubsection{IFX}

IFX is also based on a Feistel network as the FFX algorithms but supports
non uniform data. For example, part of data can be numeric and the
rest can be a string~\cite{Dworkin2013}.

\subsection{Order Revealing Encryption (ORE)}

Here the ordering is not preserved, but there exists a comparison
function which reveals the order of any two ciphertexts. Instead of
using < for comparison, this user defined comparison function is used.
It is IND-COPA secure, but has a big performance penalty. Practical
ORE schemes reveal information about the relative distance of the
plain text~\cite{10.1007/978-3-319-78556-1_3}.

\subsection{Searchable Encryption}

The ciphertext can be searched with an encrypted search term without
decrypting the ciphertext. Searchable encryption can be seen a special
case of functional encryption. Not a cipher, but another approach
is this: Additional fields are used to store keyword ciphertexts for
fuzzy queries. For each encrypted field in the database an additional
keyword field is added. On enciphering string data it generates an
keyword string which represents all sub strings of the data. After
that this string is encrypted and stored in the keyword field. During
the query it transforms the original query into combinations of terms,
and searches in the keyword field and not in the original data field~\cite{10.1109/EIDWT.2013.43,cryptoeprint:2009:251}.

\subsection{Prefix Preserving Encryption}

This is just a variant of searchable encryption: The ciphertext can
be searched by prefix in the same way like the plaintext.

\subsection{Deterministic Encryption}

The same plain text is always encrypted to the same ciphertext. AES
CBC mode is an example of deterministic encryption.

\subsection{Commutative Encryption}

The order of two encryptions and decryptions on the same plain text
is not relevant. e1(e2(P)) == e2(e1(P)), d1(d2(C)) == d2(d1(C))

\subsection{Homomorphic Encryption}

According to~\cite{DBLP:reference/crypt/2011} homomorphic encryption
is defined as ``An encryption mechanism E is called homomorphic basically
if it preserves certain algebraic structure between the plain text
space and the ciphertext space,where the encryption key is fixed.''
Homomorphic encryption makes it possible, to execute operations on
cipher text and after decryption this operations are visible on the
clear text. An example for this is RSA , where the product of two
ciphertexts is a valid encryption of the product of the corresponding
plaintexts. Fully homomorphic encryption (FHE)~\cite{cryptoeprint:2009:571,cryptoeprint:2011:277,cryptoeprint:2011:405,cryptoeprint:2011:440,DBLP:conf/eurocrypt/GentryH11,Dijk_Gentry_Halevi_Vaikuntanathan_2009} is a special kind of homomorphic encryption, where two operations, addition
and multiplication in any number and any order are supported. This
makes it possible to preserve the (algebraic) ring structure of the
plain text and by supporting these two operations any computable function
can be evaluated on encrypted values. An important application of
such an encryption scheme in the context of a database is to apply
aggregate functions like sum or average directly on encrypted data,
without ever revealing the plain text. Some homomorphic encryption
schemes exist:
\begin{itemize}
\item El Gamal Encryption Scheme~\cite[pp 308]{Vacca:2016:CCS:3098688}.
\item Pailliar Ebcryption Scheme~\cite[pp 309]{Vacca:2016:CCS:3098688}.
\item BGN (Boneh, Goh, and Nissim) Encryption Scheme~\cite[pp 311]{Vacca:2016:CCS:3098688}. Addition
and Multiplication is supported, but only one multiplication.
\item GSW (Gentry Sahai Waters) Encryption scheme (fully homomorphic).
\end{itemize}
Fully homomorphic encryption has a big performance penalty. It is
up to 1000000x slower than without encryption. Although improved,
practical use is still not possible due the limited performance~\cite{DBLP:journals/corr/GahiGE15}.

\subsection{Functional Encryption}

In contrast to the previous ciphers, this is an asymmetric encryption
scheme. Normal ciphers have no way to reveal only a particular information
about the plain text. Functional encryption gives a more fine grained
access to information. With a master key the ciphertext is still decrypted
to the plain text. Additionally to this master key a kind of sub-key
exists. With this key the ciphertext cannot decrypted, but an associated
function can performed with this key on the ciphertext, only revealing
a special property of the plain text and not the whole plain text
itself~\cite[pp 315]{Vacca:2016:CCS:3098688,Boneh:2012:FEN:2366316.2366333}.

%\newpage{}

\section{Other Solutions \label{sec:Related-Work}}

\subsection{Eperi Gateway}

Eperi Data Protection for Databases\footnote{http://eperi.de/produkte/database-encryption/}
is an approach to store only encrypted data in the database.
For processing, the data is decrypted by stored procedures so the cleartext
is in memory of DB-Server, thus providing no real end-to-end encryption.
The encryption and decryption is transparent to an application, so
the application does not have to be changed. Eperi gateway
itself provides a lot more features. It is the central point for
accessing and retrieving data from the cloud. It supports transparent
encryption/decryption for different types of data and documents. A disadvantage of this
central gateway approach is, that as all traffic to the cloud comes
and goes through the gateway. Thus, it can be a performance bottleneck and
a single point of failure~\cite{eperi}.
%\begin{figure}[ht]
%\centering
%\includegraphics[width=\columnwidth]{website_eperi-gateway-for-databases_schaubild-3}
%
%\caption{Eperi Gateway}
%
%\end{figure}

\subsection{Relational cloud}

Relational cloud\footnote{http://relationalcloud.com} is a project from
MIT to explore and enhance technology of the database as a service
(DBaaS) model in cloud computing. The vision of Relational cloud is
to provide access to all features of a DBMS without the need to manage
hardware, software and privacy.
Relational cloud consists of multiple nodes running a single database
server. Applications communicate by their standard interfaces
like JDBC. A special driver is used to connect to the front-end
to ensure data is kept private. A router is consulted by the front
end to analyze the queries and it determines the execution nodes~\cite{curino2011relational}.

\subsection{CryptDB}

CryptDB~\cite{Popa:2011:CPC:2043556.2043566} is a database management system, which can execute SQL queries
over encrypted data. It follows an SQL-aware encryption strategy and
 evaluates the query directly on the server. The client must only decrypt
the results and does not need to perform any query processing. The encryption can be completely transparent to an application,
as long as the provided client front-end is used. CryptDB uses order
preserving and homomorphic encryption.
%\begin{figure}[ht]
%\centering
%\includegraphics[width=\columnwidth]{cryptDB_overview}
%
%\caption{Architecture CryptDB\cite[pp 86]{Popa:2011:CPC:2043556.2043566} }
%\end{figure}

The goal of CryptDB is to be as secure as possible, while still providing
practical access to the database.
However, CryptDB's security was analyzed and some successful attacks were revealed:
\begin{itemize}
\item LP-optimization. 
Based on combinatorial optimization techniques, this
attack targets deterministic encryption schemes.
\item Sorting attack. 
This attack decrypts order preserving encrypted columns.
It works on dense sets, where nearly every value exists in the database.
\item Cumulative attack. 
Another attack on order preserving encrypted columns.
It works even on low-density columns by using combinatorial optimization
techniques.
\end{itemize}

%\subsubsection*{Solutions influenced by CryptDB}

According to~\cite{Tu:2013:PAQ:2535573.2488336}, there exist some
CryptDB based or at least inspired solutions:
\begin{itemize}
\item Monomi~\cite{Tu:2013:PAQ:2535573.2488336}. 
It is based on the design of CryptDB but with a focus on analytical
queries. It runs queries on encrypted data on top of the PostgreSQL
database. For the execution of complex queries, it uses the split
client/server approach (similar to~\cite{Hacigumus:2002:ESO:564691.564717}),
where part of the query is executed on the encrypted data on the server,
and another part of the query is executed on the decrypted data on
the client.
\item Microsoft's Always Encrypted SQL Server~\cite{antonopoulos2020azure}.
\item Encrypted Bigquery\footnote{https://github.com/google/encrypted-bigquery-client}. 
It is a cloud service for analysis of large data-sets from Google storage. For queries it uses an SQL dialect. Encrypted BigQuery is an extension to the Bigquery client, which is capable of client-side encryption for a subset of query types.
\item Skyhigh Networks\footnote{https://www.skyhighsecurity.com}. 
McAfee Skyhigh Security Cloud is a cloud access security broker (CASB).
This software sits between cloud service consumers and cloud service
providers to enforce security, compliance, and governance policies
for cloud applications. Among tons of other features, it supports
encryption schemes for transparent encrypting of data going through
the broker into the cloud. The Skyhigh Security Cloud supports, besides
regular symmetric encryption schemes, format-preserving and searchable
encryption schemes. Among other schemes, order preserving encryption
is supported too.
\end{itemize}

\subsection{DiCE - A Data Encryption Proxy for the Cloud}

Order-preserving encryption algorithms range from simple algorithms like Caesar encryption to secure algorithms like mOPE~\cite{popa_modular}. In order to be able to use these algorithms as easy as possible, DiCE~\cite{koppenwallner2023dice} a JDBC driver was developed\footnote{https://github.com/dicejk/dice}, that parses SQL queries as a proxy and transparently encrypts and decrypts these queries. This allows to execute many queries on an encrypted database in the cloud with (nearly) the performance as on unencrypted databases. The DiCE driver can be used with any other JDBC driver and therefore supports a variety of databases. The driver can be configured to support different encryption algorithms.

This research was motivated by our experience with large scale data stores~\cite{schikuta1998vipios} and complex applications in Grids and Clouds~\cite{mach2012generic,schikuta2004n2grid,cs745,weishaeupl2004}, and strongly motivated by our focus on Web-based workflow optimizations~\cite{schikuta2008grid,kofler2009parallel} and their respective management~\cite{stuermer2009building}.

\subsection{SecureDBaaS}

SecureDBaaS~\cite{jagadeeswaraiah2015securedbaas} differs from other solutions on the application level (e.g. DiCE~\cite{koppenwallner2023dice}),
that it does not need a proxy to store metadata. Instead all metadata
is stored encrypted on the server to avoid scalability issues. Features
of SecureDBaaS are
\begin{itemize}
\item guaranteed confidentiality by allowing concurrent SQL over encrypted data,
\item same availability, elasticity and scalability as unencrypted database as a service,
\item concurrent access from distributed clients,
\item no trusted broker or proxy required, and
\item compatible with most relational Databases. It is possible to use existing
database servers like PostgreSQL or MS SQL Server.
\end{itemize}

\subsection{CipherCloud}

CipherCloud\footnote{http://www.ciphercloud.com/} works as gateway which intercepts any traffic
between a database and its clients. In fact, it uses an enhanced JDBC
driver and supports format preserving encryption. It works with Amazon RDS as a database in the cloud~\cite{AWS_data_at_rest}.
%\begin{figure}[ht]
%\centering
%\includegraphics[width=\columnwidth]{aws_cipher_cloud_grey}
%
%\caption{Cipher Cloud}
%\end{figure}

\subsection{Voltage Secure}

Voltage Secure\footnote{https://voltage.com} provides another solution for Amazon relational database
service. Unlike CipherCloud it provides its service for applications
in the elastic cloud although with external key management. Similar to CipherCloud,
it is possible with this solution to query over encrypted data~\cite{AWS_data_at_rest}.
%\begin{figure}[ht]
%\centering
%\includegraphics[width=\columnwidth]{aws_voltage_grey}
%
%\caption{Voltage Secure}
%\end{figure}

\subsection{Perspecsys}

CloudSOC~\cite{sym_token} is a cloud access security broker (CASB). Part of it (Symantec
Cloud Data Protection \& Security) provides encryption schemes for
the data stored in the cloud. Additionally to the standard schemes,
it supports different functionality (including order) preserving encryption
schemes.

\section{Conclusion and Outlook}

Collecting and storing only the minimal required data seems obvious,
but often it is not. The GDPR makes data minimization of personal
data a principle and from a security point of view this makes absolutely
sense. Never collected data does not have to be stored, maintained
and protected. But there are enough systems, where sensible data is
definitely needed. Here it is crucial to make this classification
as sensible data explicit. Without this knowledge, it it is impossible
to protect it accordingly. Mixing sensitive with non sensitive data
is also a bad idea, because then everything has to be handled as sensitive
data and this comes with an overhead. Often it is good practice to
separate sensitive tables or even systems from non sensitives, but
this will work only if considered from the start. This report discussed
many ciphers, with different properties and different strength of
security. It is recommended to always go for the best matching cipher
regarding security requirements and needed functionality, even if
that means to use multiple different ciphers. Here, the use of the
smallest common denominator is definitely no good idea, as there is
no perfect cipher which is best for all use cases. Ciphers with specific
properties required for queries on encrypted data are available and
usable. Order preserving encryption schemes with best possible security
exist, but they are still not as secure as algorithms without additional
properties. It is important to know, that these ciphers are relatively
new and not analyzed in depth like standard algorithms as AES for
example. Of course often not optimal encryption is better than none,
but it is important to know the weaknesses of these ciphers. As a
result, is not recommended to use any non standard encryption schemes
for high classified data. The use of these ciphers also does not come
for free. There is a certain overhead by embedding the encryption
in a JDBC driver compared to using the native JDBC driver of the database
directly. The reason for this is that much more processing has to
be done before sending the query to the server. The SQL statement
has to be parsed, encrypted and sometimes rewritten before it is forwarded
to the real driver. The results of a query have to be parsed and decrypted
again. The performance results show that the response time is significant
higher, but by simulating multiple simultaneous queries the impact
on the throughput is not as dramatic. Compared with the naive approach,
by not enabling joins or indices on the encrypted data the result
is clear: The naive approach is simple not working for anything but
toy projects. Of course it can not repeated more often
that security is more than encryption, and a lot more than encrypting
the database has to be done to achieve security. Using the database
as service solutions from AWS and Azure worked like a charm. It literally
takes only 5 minutes and a database is up and running and available
for operation. Network access and speed is definitely an issue. For
example the data setup for evaluation took some time to get it into
the cloud.

%\subsection{Outlook}

Optimal order preserving encryption is here, but it has to prove itself
over the next years. As of today no high quality standard implementation
like for example for AES exist. This will hopefully change soon. The
big elephant in the room is fully homomorphic encryption, and future
will show if it is possible to develop a secure and well performing
cipher. If it is possible to achieve this, the impact goes far beyond
database encryption, because then it would be possible to directly
operate on the encrypted data and move the whole data processing in
the cloud, too. Quantum computing could be a disruptive technology
for many ciphers, which are currently believed to be secure. The adoption
of cloud computing is continuing to grow, but as more and more important
services rely on it, it seems very likely that more legislative control
will be applied to it. Privacy and security in and outside the cloud
will be more important than ever, and legislative regulation like
GDPR will become more and more relevant.

%\bibliographystyle{IEEEtran}
%\bibliography{survey}

% Generated by IEEEtran.bst, version: 1.14 (2015/08/26)

\end{document}